\documentclass[10pt,a4paper]{amsart}

\usepackage{amsthm,amssymb,amsmath,epic,eepic,float}
\usepackage{rotating,epsfig,indentfirst,array,varioref}
\usepackage{appendix,marginnote,bbm,tikz,pgf,mathtools}

\usepackage{color}

\newtheorem{Proposition}{Proposition}[section]

\newtheorem{Conj}{Conjecture}[section]

\let\emptyset\varnothing

\def\ll{\left\lgroup}
\def\rr{\right\rgroup}

  \def\rrho{\tt R}
\def\ssigma{\tt C}

\def\leq{\leqslant}
\def\geq{\geqslant}

\def\pp{p^{\prime}}

\def\i{\iota}

\restylefloat{figure}

\newcommand{\cB}{\mathcal{B}}

\newcommand{\cE}{\mathcal{E}}

\newcommand{\cF}{\mathcal{F}}

\newcommand{\cH}{\mathcal{H}}

\newcommand{\cM}{\mathcal{M}}
\newcommand{\cN}{\mathcal{N}}
\newcommand{\cO}{\mathcal{O}}

\newcommand{\cS}{\mathcal{S}}

\newcommand{\cV}{\mathcal{V}}
\newcommand{\cW}{\mathcal{W}}

\newcommand{\RR}{\mathbb{R}}

\newcommand{\apos}{\alpha_{+}}
\newcommand{\aneg}{\alpha_{-}}

\newtheorem{ca}{Figure}

\def\ll{ \left\lgroup}
\def\rr{\right\rgroup}

\newcommand{\checkedcell}{{\makebox[0pt][l]{$\square$}\raisebox{.15ex}{\hspace{0.1em}$\checkmark$}}}

\begin{document}

\title[AGT, Burge pairs and minimal models]{
AGT, Burge pairs and minimal models}

\author[Bershtein and Foda]{M Bershtein \!$^{ {\scriptstyle \mathbbm{L}} }$ and 
                                 O Foda \!$^{ {\scriptstyle \mathbbm{M}} }$}

\address{
\!\!\!\!\!\!\!\!\!$^{ {\scriptstyle \mathbbm{L}} }$ Landau Institute for Theoretical Physics, Chernogolovka, Russia
\newline
Institute for Information Transmission Problems, Moscow, Russia
\newline 
National Research University Higher School of Economics,
International Laboratory of Representation Theory and Mathematical Physics,
\newline 
Independent University of Moscow, Moscow, Russia
\newline
$^{ {\scriptstyle \mathbbm{M}} }$ Mathematics and Statistics, University of Melbourne, Parkville, VIC 3010, Australia
}

\email{mbersht@gmail.com, omar.foda@unimelb.edu.au}

\keywords{
AGT correspondence. 
Burge pairs.
Minimal Virasoro conformal field theories.
Conformal blocks.
}

\begin{abstract}
We consider the AGT correspondence in the context of the conformal field theory $\cM^{\, p, \pp}$ $\otimes$ 
$\cM^{\cH}$, where $\cM^{\, p, \pp}$ is the minimal model based on the Virasoro algebra $\cV^{\, p, \pp}$ 
labeled by two co-prime integers $\{p, \pp\}$, $1 < p < \pp$, and $\cM^{\cH}$ is the free boson theory 
based on the Heisenberg algebra $\cH$. Using Nekrasov's instanton partition functions without modification 
to compute conformal blocks in $\cM^{\, p, \pp}$ $\otimes$ $\cM^{\cH}$ leads to ill-defined or incorrect 
expressions.

Let $\cB^{\, p, \pp, \cH}_n$ be a conformal block in $\cM^{\, p, \pp}$ $\otimes$ $\cM^{\cH}$, with $n$ 
consecutive channels 
$\chi_{\i}$, $\i = 1, \cdots, n$, and let $\chi_{\i}$ carry states from 
$\cH^{p, \pp}_{r_{\i}, s_{\i}}$ $\otimes$ $\cF$, where 
$\cH^{p, \pp}_{r_{\i}, s_{\i}}$ is an irreducible highest-weight $\cV^{\, p, \pp}$-representation, labeled 
by two integers $\{r_{\i}, s_{\i}\}$, $0 < r_{\i} < p$, $0 < s_{\i} < \pp$, and $\cF$ is the Fock space of
$\cH$. 

We show that restricting the states that flow in $\chi_{\i}$ to states labeled by 
a partition pair $\{Y_1^{\i}, Y_2^{\i}\}$ such that 
$Y^{\i}_{2, \rrho} - Y^{\i}_{1, \rrho +       s_{\i} - 1} \geq 1 -     r_{\i}$, and
$Y^{\i}_{1, \rrho} - Y^{\i}_{2, \rrho + \pp - s_{\i} - 1} \geq 1 - p + r_{\i}$, 
where $Y^{\i}_{i, \rrho}$ is row-$\rrho$ of $Y^{\i}_i, i \in \{1, 2\}$, 
we obtain a well-defined expression that we identify with $\cB^{\, p, \pp, \cH}_n$. 
We check the correctness of this expression for 
{\bf 1.} Any 1-point $\cB^{\, p, \pp, \cH}_1$ on the torus, when the operator insertion is the identity, 
and 
{\bf 2.} The 6-point $\cB^{\, 3,   4, \cH}_3$ on the sphere that involves six Ising magnetic operators. 
\end{abstract}
\maketitle
%
\section{Introduction}
\label{introduction}
\subsection{AGT in generic models}
Consider the two-dimensional conformal field theory 
$\cM^{\, \textit{gen}, \cH}$ $=$ $\cM^{\, \textit{gen}}$ $\otimes$ $\cM^{\cH}$,
based on the algebra
$\cV^{\, \textit{gen}, \cH}$ $=$ $\cV^{\, \textit{gen}}$ $\oplus$ $\cH$, where 
$\cM^{\, \textit{gen}}$ is a generic model with a chiral spectrum that spans infinitely-many 
infinite-dimensional irreducible highest-weight $\cV^{\, \textit{gen}}$-representations
\footnote{\ Only the chiral sector of a conformal field theory is discussed in this work, 
and this is implied in the sequel. We abbreviate 
{\it \lq the AGT correspondence\rq\ } to {\it \lq AGT\rq\ }, and
{\it \lq irreducible highest-weight representation\rq\ } to
{\it \lq irrep\rq}, which in this work is always infinite-dimensional.},
$\cM^{\cH}$ is the conformal field theory of a free boson that takes values in $\RR$,
$\cV^{\, \textit{gen}}$ is the Virasoro algebra of generic central charge $c_{\textit{gen}}$
\footnote{\ By generic Virasoro central charge $c_{\textit{gen}}$ we specifically mean 
$c_{\textit{gen}}$ $\neq$ $c_{p, \pp}$, where $c_{p, \pp}$ is the Virasoro central charge 
of the minimal model $\cM^{p, \pp}$.}, 
and $\cH$ is the Heisenberg algebra. The Virasoro central charge of $\cM^{\cH}$ is $c_{\cH}$ $= 1$. 
The AGT correspondence of Alday, Gaiotto and Tachikawa \cite{alday.gaiotto.tachikawa} identifies conformal blocks in $\cM^{\, 
\textit{gen}, \cH}$ \cite{belavin.polyakov.zamolodchikov} with instanton partition functions 
in four-dimensional $\cN \! = \! 2$ supersymmetric quiver gauge theories \cite{nekrasov}. 
Conjectured in \cite{alday.gaiotto.tachikawa}, AGT was proven for 
$c_{\textit{gen}} \! = \! 1$ in \cite{mironov.morozov.shakirov}, and for all $c_{\textit{gen}}$ 
in \cite{alba.fateev.litvinov.tarnopolskiy} for conformal blocks with non-degenerate external 
primary fields. 

\subsection{AGT in minimal models} In this note, we consider AGT in the context of 
$\cM^{\, p, \pp, \cH}$ $=$ $\cM^{\, p, \pp}$ $\otimes$ $\cM^{\cH}$, based on the algebra
$\cV^{\, p, \pp, \cH}$ $=$ $\cV^{\, p, \pp}$ $\oplus$ $\cH$, 
where 
$\cM^{\, p, \pp}$ is the minimal conformal field theory with a chiral spectrum that spans
finitely-many $\cV^{\, p, \pp}$ irreps, and $\cV^{\, p, \pp}$ is the Virasoro algebra 
labeled by two co-prime integers $\{p, \pp\}$, $0 < p < \pp$, of central charge $c_{p, \pp}$, 

\begin{equation}
\label{c.p.p.prime}
c_{p, \pp} = 1 - 6 \ll \ll \frac{p}{\pp} \rr^{\frac12}  
                     - \ll \frac{\pp}{p} \rr^{\frac12} \rr^2 
\end{equation}

\noindent Let $\cB^{\, p, \pp, \cH}_n$, $\cB^{\, p, \pp}_n$ and $\cB^{\, \cH}$ be conformal 
blocks with $n$ consecutive channels 
\footnote{\ Only linear conformal blocks, as in Figure {\bf 3}, 
are considered in this work. Our notation is such that an $n$-channel conformal block 
$\cB^{\, \textit{indices}}_n$, is the expectation value of $(n+3)$ vertex operators
$\cO^{\, \textit{same indices}}_{\i} (z_{\i})$, ${\i} = 0, \cdots, (n+2)$, in 
$\cM^{\, \textit{same indices}}$ on a Riemann surface $\cS$, and $z_{\i} \in \cS$.}.
We wish to compute any 
$\cB^{\, p, \pp, \cH}_n$ of vertex operators 
$\cO^{\, p, \pp, \cH}_{\i} (z_{\i})$ $=$ $\cO^{\, p, \pp}_{\i} (z_{\i})$ $\times$ $\cO^{\, \cH}_{\i} (z_{\i})$, 
$\i = 0, \cdots, n+2$. 
Since any $\cB^{\, p, \pp, \cH}_n$ factorizes to $\cB^{\, p, \pp}_n$ $\times$ $\cB^{\, \cH}_n$ and an 
explicit expression for the $\cM^{\, \cH}$-factor $\cB^{\, \cH}_n$ is known
\footnote{\ See, for example, equation (1.9) in \cite{alba.fateev.litvinov.tarnopolskiy}.}, 
computing $\cB^{\, p, \pp, \cH}_n$ is equivalent to computing 
its $\cM^{\, p, \pp}$-factor $\cB^{\, p, \pp}_n$ which is typically what we want. 

\subsection{Zeros in denominators and deformations} 
\label{deformations}
Applying AGT to minimal model conformal blocks 
without modification leads to ill-defined expressions, as will be explained in detail below. In particular, 
setting the parameters that appear in Nekrasov's partition functions to minimal model values leads to zeros in 
the denominators of the summands. Following \cite{santachiara.0}, one can make the summands well-defined using 
suitable deformations of the parameters. Doing that, one finds whenever a denominator is zero in the limit of 
removing the deformations, the corresponding numerator is also zero in such a way that and that limit is 
well-defined. This is in agreement with \cite{zamolodchikov}, where arguments were given to the effect that, 
analytically continuing the conformal blocks in the conformal dimensions of the primary states that flow in 
each channel, the only singularities are poles and the sum of all residues is zero. This is the approach that 
was followed, albeit without discussion, in an earlier work on AGT in minimal models \cite{santachiara.2}. 

\subsection{Zeros in denominators and restrictions}
\label{restrictions}
In this note, we follow a different approach from that discussed in subsection {\bf \ref{deformations}}. Our 
idea is that the zeros in the denominators of Nekrasov's 
partition functions are due to including null states that should not be included. We avoid this by restricting 
the summations over Young diagrams that appear in Nekrasov partition functions to avoid these null states. We 
make the summands well-defined by restricting the partition pairs that label the summed-over states to exclude 
the summands with poles. 
To compute $\cB^{\, p, \pp, \cH}_n$, the summations that label the factors in Nekrasov's instanton partition 
functions must be restricted to avoid ill-defined or incorrect expressions for $\cB^{\, p, \pp, \cH}_n$, and 
consequently for its $\cM^{\, p, \pp}$-factor $\cB^{\, p, \pp}_n$. Our approach allows us to characterise the 
Young diagrams that label the summands that do contribute to $\cB^{\, p, \pp, \cH}_n$.

\subsection{Unrestricted partition pairs} 
\label{unrestricted.partition.pairs}
The AGT expression for a linear conformal block $\cB^{\, \textit{gen}, \cH}_n$, that has $n$ consecutive 
channels $\chi_{\i}$, $\i = 1, \cdots, n$, is an $n$-fold sum
\footnote{\ 
The partition pairs $\vec Y^0$ and $\vec Y^{n+1}$ are trivial, that is they consist of empty partitions,
and no summation is performed on them.}, 

\begin{equation}
\cB^{\, \textit{gen}, \cH}_n=\sum_{\vec Y^{1}, \cdots, \vec Y^{n}}
\prod_{\i = 1}^{n+1} q_{\i}^{| \vec Y^{\i } |} 
Z_{bb}^{\i} \ll \vec a^{\i - 1}, \vec Y^{\i - 1}\ | \ \mu^{\i } \ | \ \vec a^{\i}, \vec Y^{\i} \rr,
\end{equation}

\noindent where the summand is a product of $(n+1)$ factors 
$q_{\i}^{| \vec Y^{\i} |} Z_{bb}^{\i} [\vec a^{\i - 1}, \vec Y^{\i - 1}\ | \ \mu^{\i} \ | \ \vec a^{\i}, \vec Y^{\i}]$,
$\i = 1, \cdots, n+1$, that will be defined in section {\bf 2}. Each factor $Z_{bb}^{\i}$ is a rational function that 
depends on two pairs of {\it \lq unrestricted\rq} Young diagrams 
$\{Y^{\i - 1}_1, Y^{\i - 1}_2\}$ and $\{Y^{\i}_1, Y^{\i}_2\}$. In other words, there are 
no conditions on these Young diagrams and all possible pairs are allowed. The denominator $z_{den}^{\i}$ of 
$Z_{bb}^{\i}$ is a product of the norms of the states that flow in the preceding channel $\chi^{\i - 1}$ and the 
subsequent channel $\chi^{\i}$.
Since $Z_{bb}^{\i}$ is labeled by unrestricted partition pairs, and the sums are over all possible unrestricted 
pairs, the states that flow in each channel belong to a Verma module of $\cV^{\, \textit{gen},    \cH}$.

Applying AGT without modification to $\cM^{\, p, \pp, \cH}$, one includes zero-norm states in the summation, 
and thereby includes states in a Verma module rather than in an irrep of 
$\cV^{\, p, \pp, \cH}$. This leads to summands in the instanton partition function with zero denominators. 
Further, as show, whenever a denominator in a summand vanishes, the corresponding numerator vanishes as well 
and one ends with ill-defined expressions

\subsection{Restricted partition pairs} 
\label{restricted.partition.pairs}
In this note, we consider $\cB^{p, \pp, \cH}_n$ as an instanton partition $Z_{Nek}$ that consists of building 
block partition functions $Z^{\i}_{bb}$ that has a numerator $z^{\i}_{num}$ and a denominator $z^{\i}_{den}$,
$\i = 1, \cdots, n+1$. $Z^{\i}_{bb}$ connects two channels $\chi_{\i - 1}$ and $\chi_{\i}$. The denominator 
$z^{\i}$ of $Z^{\i}_{bb}$ is a product of two factors $[z^{\i - 1}_{norm}]^{1/2}$ and $[z^{\i}_{norm}]^{1/2}$ 
that account for the norms of the states that flow in the channels $\chi_{\i -1}$ and $\chi{\i}$, respectively.
We characterise the zeros in these denominators that lead to ill-defined expressions for $\cB^{\, p, \pp, \cH}_n$. 
If channel $\chi_{\i}$, $\i = 1, \cdots, n$, carries states that belong to an irreducible highest weight Virasoro 
representation that flows is $\cH^{p, \pp}_{r_{\i}, s_{\i}}$, we attribute these zeros to the flow of null 
states that do not belong to  $\cH^{p, \pp}_{r_{\i}, s_{\i}}$, and eliminate these zeros by restricting the 
partition pairs that appear in Nekrasov's original expressions to partition pairs $\{Y_1, Y_2\}$, that satisfy 
the conditions  

\begin{equation}
\label{Burge.conditions}
\boxed{
Y^{\i}_{2, \rrho} - Y^{\i}_{1, \rrho +     s_{\i} - 1} \geq 1 -       r_{\i}, 
\quad
Y^{\i}_{1, \rrho} - Y^{\i}_{2, \rrho + p - s_{\i} - 1} \geq 1 - \pp + r_{\i}
}
\end{equation}

\noindent where $Y^{\i}_{i, \rrho}$ is row-$\rrho$ of $Y^{\i}_i$, $i$ $\in$ $\{1, 2\}$.

\subsection{Burge pairs} Partition pairs that satisfy conditions (\ref{Burge.conditions}) 
were first studied in \cite{burge} and appeared more recently in 
\cite{foda.lee.welsh, feigin.feigin.jimbo.miwa.mukhin}. In this work, we refer to them as 
{\it Burge pairs}, and show that when used to restrict AGT to compute $\cB^{\, p, \pp, \cH}_n$, 
that is when we sum over Burge pairs rather than on all possible partition pairs, 

\begin{equation}
\label{restricted.sum}
\cB^{\, p, \pp, \cH}_n =\sum^{\prime}_{\vec Y^1, \cdots, \vec Y^n}
\prod_{\i = 1}^{n+1} q_{\i}^{| \vec Y^{\i } |} 
Z_{bb}^{\i} \ll \vec a^{\i - 1}, \vec Y^{\i - 1}\ | \ \mu^{\i } \ | \ \vec a^{\i}, \vec Y^{\i} \rr
\end{equation}

\noindent where $\sum^{\prime}$ indicates that the sum is restricted to partition pairs that satisfy the 
Burge conditions (\ref{Burge.conditions}), we obtain well-defined expressions. We check these expressions 
in two cases
{\bf 1.} Any 1-point $\cB^{\, p, \pp, \cH}_1$ on the torus, when the operator insertion is the identity, 
and 
{\bf 2.} The 6-point $\cB^{\, 3,   4, \cH}_3$, when all operator insertions involve Ising magnetic
operators. We also give arguments why we expect this identification to be correct.

\subsection{Outline of contents}
In section {\bf 2}, we recall basic facts related to Nekrasov's instanton partition functions. 
In {\bf 3}, we recall the AGT parametrisation of  $\cM^{\, \textit{gen}, \cH}$, 
the choice of parameters that allows us to obtain $\cM^{\, p, \pp, \cH}$, 
then show how the unrestricted instanton partition functions give the wrong answer in 
the case of $\cB^{\, p, \pp, \cH}_1$ on the torus.
In {\bf 4}, we use the requirement that the summands remain well-defined to characterise 
the partition pairs that label them. We identify these partition functions 
with $\cB^{\, p, \pp, \cH}_n$ 
In {\bf 5}, we study the vanishing of the numerator, and show that whenever the denominator 
of a summand vanishes, then the numerator also vanishes. 
In {\bf 6}, we check the correctness of our expressions in the two cases listed above. 
In {\bf 7}, we use results from 
\cite{alba.fateev.litvinov.tarnopolskiy,foda.lee.welsh, feigin.feigin.jimbo.miwa.mukhin},
Proposition {\bf \ref{proposition.01}} in section {\bf 4}, and  
Conjecture {\bf \ref{conjecture.01}} in section {\bf 7}, to explain why the restriction 
to Burge pairs  produces conformal blocks in $\cM^{\, p, \pp, \cH}$. 
Because we use Conjecture 7.1, this explanation is not a proof.
In {\bf 8}, we extend of our results to conformal blocks in $\cM^{gen, \cH}$, with 
degenerate intermediate Virasoro representations, and
in {\bf 9}, we collect a number of remarks that include 
{\bf {$\mathfrak{1}$}.} a conjectural generalization to the $W_N$ conformal blocks, and 
{\bf {$\mathfrak{2}$}.} a geometric interpretation of the summation over Burge pairs as 
a summation over {\it isolated} torus fixed points on the instanton moduli space.

\vfill
\newpage

\section{The instanton partition function}
\label{nekrasov.partition.functions}

\subsection{Partitions}
\label{partitions}
A partition $\pi$ of an integer $|\pi|$ is a set of non-negative integers $\{\pi_1, \pi_2,$
$ \cdots, \pi_p\}$, where $p$ is the number of parts, $\pi_i \geq \pi_{i+1}$, and 
$\sum_{i=1}^{p} \pi_i= |\pi|$. $\pi$ is represented as a Young diagram $Y$, 
which is a set of $p$ rows $\{Y_1, Y_2, \cdots, Y_p\}$, such that row-$i$ has 
$Y_i = \pi_i$ cells 
\footnote{\ We use $Y_i$ for row-$i$ as well as for the number of cells in that row.}, 
$Y_i \geq Y_{i+1}$, and $| Y | = \sum_i Y_i$ $= |\pi|$. We use $Y_i^{\intercal}$ for the 
transpose of $Y_i$, and define $Y_{i, \rrho}^+ = Y_{i, \rrho} + 1$. 

We use $\square$ for a cell in a Young diagram $Y$, which is a square in the south-east 
quadrant of the plane, with coordinates $\{\rrho, \ssigma\}$, such that 
$\rrho$   is the row-number,    counted from top  to bottom, and 
$\ssigma$ is the column number, counted from left to right. 
We define $A^+_{\square, Y_i} = A_{\square, Y_i} + 1$, where 
$A_{\square, Y_i}$ is the arm of $\square$ in $Y_i$, 
that is, the number of cells in the same row as, but to the right of $\square$ in $Y_i$,
and
$L_{\square, W_j}$ to be the leg of $\square$ with respect its position in $W_j$, 
that is the number of cells in the same column as, but below $\square$ in $Y_i$. 

%
\begin{center}
\begin{minipage}{5.0in}
\setlength{\unitlength}{0.001cm}
\renewcommand{\dashlinestretch}{30}
\begin{picture}(4800, 3200)(-4000, 1400)
%
\thicklines
%
\path(0000,3600)(3000,3600)
\path(0000,3000)(3000,3000)
\path(0000,2400)(2400,2400)
\path(0000,1800)(2400,1800)
\put(-0500,3150){$Y_1$}
\put(-0500,2550){$Y_2$}
\put(-0500,1950){$Y_3$}
%
\path(0000,3600)(0000,1800)
\path(0600,3600)(0600,1800)
\path(1200,3600)(1200,1800)
\path(1800,3600)(1800,1800)
\path(2400,3600)(2400,1800)
\path(3000,3600)(3000,3000)
\put(0050,3800){$Y^{\intercal}_1$}
\put(0650,3800){$Y^{\intercal}_2$}
\put(1250,3800){$Y^{\intercal}_3$}
\put(1850,3800){$Y^{\intercal}_4$}
\put(2450,3800){$Y^{\intercal}_5$}
\put(0750,2600){$\checkmark$}
\end{picture}
\begin{ca}
\label{1.Young.diagram}
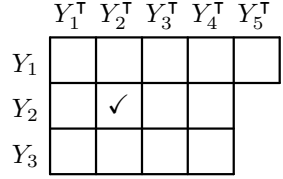
The Young diagram $Y$ of $5 +$$ 4 +$$ 4$.
The rows are numbered from top to bottom. 
The rows of the transpose $Y^{\intercal}$ that represents $3 +$$ 3 +$$ 3 +$$ 3 +$$ 1$, 
which are the columns of $Y$, are numbered from left to right. 
From the viewpoint of $Y$, the marked cell $\checkedcell$ has 
$A^{ }_{\checkedcell, Y} = 2$, 
$A^{+}_{\checkedcell, Y} = 3$, 
and 
$L_{\checkedcell, Y} = 1$.
From the viewpoint of $Y^{\intercal}$, $\checkedcell$ has
$A^{ }_{\checkedcell, Y^{\intercal}} = 1$, 
$A^{+}_{\checkedcell, Y^{\intercal}} = 2$, 
and
$L_{\checkedcell, Y^{\intercal}} = 2$.
\end{ca}
\end{minipage}
\end{center}
\bigskip

\subsection{Partition pairs}
The AGT representation of $\cB^{\, p, \pp, \cH}$ involves a multi-sum over internal states labeled 
by $n+2$ partition pairs $\vec Y^{\i}$, $\i = 0, 1, \cdots, n, n+1$, where $\vec Y^{\i}$ is a pair 
of Young diagrams, $\{ Y_1^{\i }, Y_2^{\i } \}$, and
$|\vec Y^{\i }|$ $=$ $|Y_1^{\i}| + |Y_2^{\i}|$ is the total number of cells in $\vec Y^{\i}$.
The pairs $\{ Y^{\i }_1, Y^{\i }_2 \}$, $\i \in \{1, \cdots, n\}$, are non-empty Young diagrams, 
while $\{ Y^{\i }_1, Y^{\i }_2 \}$, $\i \in \{0, n+1\}$ are empty 
\footnote{\ We work in terms of $n+2$ linearly-ordered partition pairs. Since we consider conformal 
blocks of primary fields, the initial and final pairs are always empty, but we prefer to work in terms 
of $n+2$ rather than $n$ non-empty pairs to make the notation in the sequel more uniform.}, 
$\vec Y^{(0)} = \vec Y^{(n+1)} = {\vec \emptyset}$, 
where $\vec \emptyset$ is a pair of empty Young diagrams. 

%
\begin{center}
\begin{minipage}{5.0in}
\setlength{\unitlength}{0.001cm}
\renewcommand{\dashlinestretch}{30}
\begin{picture}(4800, 3200)(-2000, 1400)
\thicklines
%
\path(0000,1800)(2400,1800)
\path(0000,2400)(2400,2400)
\path(0000,3000)(3000,3000)
\path(0000,3600)(3000,3600)
%
\path(0000,3600)(0000,1800)
\path(0600,3600)(0600,1800)
\path(1200,3600)(1200,1800)
\path(1800,3600)(1800,1800)
\path(2400,3600)(2400,1800)
\path(3000,3600)(3000,3000)
%
\path(3900,3600)(6300,3600)
\path(3900,3000)(6300,3000)
\path(3900,2400)(4500,2400)
\path(3900,1800)(4500,1800)
%
\path(3900,3600)(3900,1800)
\path(4500,3600)(4500,1800)
\path(5100,3600)(5100,3000)
\path(5700,3600)(5700,3000)
\path(6300,3600)(6300,3000)
\put(1400,2600){$\checkmark$}
\put(5300,2600){$\checkmark$}
\end{picture}
\begin{ca}
\label{2.Young.diagrams}
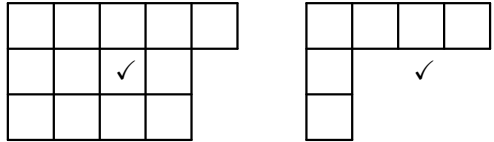
A partition pair $\{Y_1, Y_2\}$. $Y_1$ is on the left, $Y_2$ is on the right. The cell $\checkedcell$
has coordinates $(2, 3)$, $\checkedcell$ $\in Y_1$, but $\checkedcell$ $\not \in Y_2$. 
It has
$A^{ }_{\checkedcell, Y_1} = 1$,
$A^{+}_{\checkedcell, Y_1} = 2$,
$L_{\checkedcell, Y_1} = 1$,
as well as
$A^{ }_{\checkedcell, Y_2} = -2$,
$A^{+}_{\checkedcell, Y_2} = -1$,
and
$L_{\checkedcell, Y^{\intercal}} = -1$.
\end{ca}
\end{minipage}
\end{center}
\bigskip

\subsection{A decomposition of the instanton partition function}
\label{decomposition.of.the.instanton.partition.function}
Consider the four-dimen\-sional ${\mathcal N} = 2$ supersymmetric linear quiver gauge theory with 
a gauge group $\prod_{\i=1}^{n+1} U(2)_{\i}$, that is $(n+1)$ copies of $U(2)$ \cite{nekrasov}. 
The instanton partition function of this theory can be written in terms of \lq building block\rq\ 
partition functions $Z^{\i}_{bb}$, $\i = 1, \cdots, n+1$, as follows

\begin{equation}
\label{z.nek}
Z_{Nek} = 
\sum_{\vec Y^{1}, \cdots, \vec Y^{n}}
\prod_{\i = 1}^{n+1} q_{\i}^{| \vec Y^{\i } |} 
Z_{bb}^{\i} \ll \vec a^{\i - 1}, \vec Y^{\i - 1}\ | \ \mu^{\i } \ | \ \vec a^{\i}, \vec Y^{\i} \rr,
\end{equation}

\noindent where $q_{\i}$ is an indeterminate. In gauge theory, $q_{\i} = e^{2 \pi i \tau_{\i}}$, 
where $\tau_{\i}$ is the complexified coupling constant of $U(2)_{\i}$. In conformal field 
theory, it is a rational function of the positions $z_{\i}$, $\i = 0, 1, \cdots, n + 2$, of 
the vertex operators $\cO_{\i}$, whose expectation value is the conformal block, on the Riemann 
surface $\cS$ that the conformal field theory is defined on. $Z_{bb}^{\i}$ is defined in subsection 
{\bf \ref{the.building.block}}. 

The decomposition of the instanton partition function in (\ref{z.nek}) follows that in \cite{kanno} 
and mirrors the decomposition of conformal blocks on a sphere, represented as a {\it comb diagram} 
in Figure {\bf 3}. 

%
\begin{center}
\begin{minipage}{5.0in}
\setlength{\unitlength}{0.001cm}
\renewcommand{\dashlinestretch}{30}
\begin{picture}(4800, 2400)(-3000, -0500)
%
\thicklines
%
\path(0000,0000)(7200,0000)
\put(- 700,-100){$\cO_0$}
\put( 7500,-100){$\cO_6$}
\path(1200,0000)(1200,1200)
\path(2400,0000)(2400,1200)
\path(3600,0000)(3600,1200)
\path(4800,0000)(4800,1200)
\path(6000,0000)(6000,1200)
\put(1050,1400){$\cO_1$}
\put(2250,1400){$\cO_2$}
\put(3450,1400){$\cO_3$}
\put(4650,1400){$\cO_4$}
\put(5850,1400){$\cO_5$}
\put(1700,-0400){$\chi_1$}
\put(2900,-0400){$\chi_2$}
\put(4100,-0400){$\chi_3$}
\put(5300,-0400){$\chi_4$}
\end{picture}
\begin{ca}
\label{The.comb.diagram}
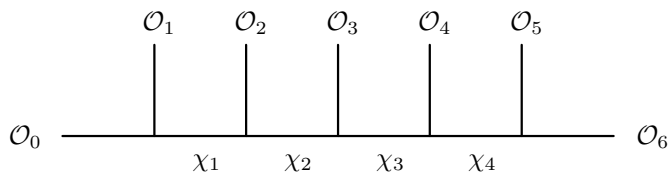
The comb diagram of a 4-channel conformal block that corresponds to a linear quiver. 
It consists of 
an initial state that corresponds to a vertex operator $\cO_0$ on the left, 
five vertex operator insertions $\cO_1, \cdots, \cO_5$, and 
a  final   state that corresponds to a vertex operator $\cO_6$ on the right.
$\cO_{\i}$ is placed at $z_{\i}$, where $z_0$, $z_{n+1}$ and $z_{n+2}$ are set $0, 1$ 
and $\infty$, respectively. In this example, $n=4$.

\end{ca}
\end{minipage}
\end{center}
\bigskip

\subsection{The building block of the instanton partition function}
\label{the.building.block}
$Z_{bb}$ is 
\begin{equation}
\label{z.bb}
Z_{bb}  \ll \vec a, \vec Y\ | \ \mu \ | \ \vec b, \vec W \rr = 
\frac{
z_{num} \ll \vec a, \vec Y\ | \ \mu \ | \ \vec b, \vec W \rr
}
{
z_{den} \ll \vec a, \vec Y\ |           \ \vec b, \vec W \rr
}, 
\end{equation}

The parameters that appear in $Z_{bb}$ are as follows.

\subsubsection{The 2-component vector $\vec a^{\i} = \{ a^{\i}, - a^{\i} \}$} 
In gauge theory, $a^{\i}$ is the expectation value of the vector multiplet in the adjoint 
representation of the gauge group $U(2)_{\i}$.
In conformal field theory, $a^{\i}$ is the charge of the highest weight of the Virasoro irrep 
that flows in channel $\chi_{\i}$ in the conformal block under consideration. 

\subsubsection{The partition pairs $\vec Y$ and $\vec W$}
In gauge theory, each partition pair $\vec Y^{\i}$ $= \{ Y_1^{\i}, Y_2^{\i}\}$ labels the fixed 
localization points in the instanton moduli space of $U(2)_{\i}$.
In conformal field theory, they label the states that flow in channel $\chi_{\i}$ in the 
corresponding conformal block. In (\ref{z.bb}), $\vec Y$ and $\vec W$ are attached to the 
line segments on the left and the right of a given vertex, respectively.

\subsubsection{The scalar $\mu^{\i}$}
In gauge theory, $\mu^{\i}$ is the mass parameter of the bi-fundamental matter field that interpolates 
the gauge groups $U(2)_{\i}$ and $U(2)_{\i + 1}$. 
In conformal field theory, $\mu^{\i}$ is the charge of the vertex operator that connects channels 
$\chi_{\i}$ and $\chi_{\i + 1}$. In the following, we study the structure of the right hand side 
of (\ref{z.bb}).

\subsubsection{The denominator} 
\label{the.denominator}
\begin{equation}
\label{z.den}
z_{den} \ll \vec a, \vec Y \ | \ \vec b, \vec W \rr
= 
\ll 
z_{norm} \ll \vec a, \vec Y \rr \ 
z_{norm} \ll \vec b, \vec W \rr 
\rr^{\frac{1}{2}},
\end{equation}

\noindent where
\begin{equation}
\label{z.norm}
z_{norm} \ll \vec a, \vec Y \rr = z_{num} \ll \vec a, \vec Y \ | \  0 \ | \ \vec a, \vec Y \rr 
\end{equation}

\noindent In gauge theory, $z_{norm}$ is a normalization factor related to the contribution of 
the vector multiplets that the bi-fundamental couples to. 
In conformal field theory, it accounts for the norms of the states that propagate into and out 
of the vertex operator insertion in $Z_{bb}$. 

\subsubsection{The numerator} 
\label{the.numerator}
\begin{multline}
\label{z.num}
z_{num} \ll \vec a, \vec Y\  | \ \mu \ | \ \vec b, \vec W \rr = 
\\
\prod_{i, j = 1}^2 
\prod_{\square \in Y_i}
\ll
E[a_i - b_j, Y_i, W_j, \square] - \mu
\rr
\prod_{\blacksquare \in W_j}
\ll
\epsilon_1 + \epsilon_2 - E[b_j - a_i, W_j, Y_i, \blacksquare] - \mu
\rr,
\end{multline}
\noindent where the elementary function $E[x, Y_i, W_j, \square]$ is defined as

\begin{equation}
\label{E.generic.1}
E[x, Y_i, W_j, \square] = x 
+ A^{+}_{\square, Y_i} \epsilon_2 
- L^{ }_{\square, W_j} \epsilon_1, 
\end{equation}
\noindent $x$ is an indeterminate, and $\{\epsilon_1, \epsilon_2\}$ are Nekrasov's deformation 
parameters, which are generally complex. 
In gauge theory, $z_{num}$ is the contribution of a bi-fundamental multiplet in $U(2)_{\i}$
and $U(2)_{\i + 1}$. 
In conformal field theory, it is the contribution of the vertex operator insertion that inputs 
a charge $\mu$ into the conformal block into $Z_{bb}$. 

\subsubsection{Remark} One can think of $z_{num}$ as the basic object in $U(2)$ AGT theory 
and in this paper, and all other objects can be written in terms of special cases of it. 

\subsubsection{Normalisation}

Consider the special case where the Virasoro part of the vertex operator in $Z_{bb}$ is the 
identity, that is $\{r, s\}$ $= \{1, 1\}$, and consequently $\mu = 0$ 
\footnote{\ 
See subsection {\bf \ref{notation.02}}.}. 
$Z_{bb}$ is defined combinatorially and does not necessarily vanish when the fusion rules
are not satisfied. To ensure that the fusion rules are satisfied, we set $\vec a = \vec b$.

Setting $\mu = 0$ and $\vec a = \vec b$ ensures that the Virasoro part of the vertex operator 
insertion is the identity operator. However, one can show that, in this case, the Heisenberg 
part of the vertex operator is an exponential of the creation part of the free boson field 
\cite{alba.fateev.litvinov.tarnopolskiy}, which in general contributes to a difference between 
$\vec Y$ and $\vec W$, and therefore we do not necessarily have $\vec Y = \vec W$. 
Setting $\vec Y = \vec W$, we pick up the contribution of the trivial part of the exponential, 
that is the identity, and $Z_{bb}$ reduces to

\begin{equation}
\label{z.bb.normalization}
Z_{bb}  \ll \vec a, \vec Y\ | \ 0 \ | \ \vec a, \vec Y \rr =
\frac{
z_{num} \ll \vec a, \vec Y\ | \ 0 \ | \ \vec a, \vec Y \rr
}
{
z_{den} \ll \vec a, \vec Y\  |        \ \vec a, \vec Y \rr
} 
=
1
\end{equation}

\noindent Equation (\ref{z.bb.normalization}) is relevant to computing 1-point conformal blocks 
of the identity operator on the torus in subsections {\bf \ref{unmodified.torus}} and {\bf \ref{modified.torus}}.

\section{Unrestricted instanton partition functions for $\cM^{\, p, \pp, \cH}$}
\label{unmodified.agt}

\subsection{AGT parameterisation. Generic models} 
\label{parameterisation.generic.models}

A generic model is a conformal field theory characterised by a central charge $c_{\textit{gen}}$ 
that we parametrise as 

\begin{equation}
\label{central.charge.generic}
c_{\textit{gen}} = 1 + 6 \ll b_{\textit{gen}} + \frac{1}{b_{\textit{gen}}} \rr^2,
\quad
b_{\textit{gen}} = \ll \frac{\epsilon_2}{\epsilon_1} \rr^{\frac{1}{2}},
\end{equation}

In the Coulomb gas approach to computing conformal blocks in generic models, the screening charges 
$\{ \beta_{+}, \beta_{-}\}$, and the background charge, $- 2 \beta_0$, satisfy 
\footnote{\ We use $\beta_{+}, \beta_{-}$, $-2\beta_0$ for generic model charges and reserve 
$\apos$, $\aneg$ and $-2 \alpha_0$ for the corresponding minimal model charges.
We use $b_{\textit{gen}}$ and $a_{p, \pp}$ for the parameters used to describe the generic and
minimal models central charges respectively, since $a$ and $b$ are used for other purposes in
the sequel.} 

\begin{equation}
\label{screening.charges.generic}
\beta_{+} = b_{\textit{gen}}, 
\quad
\beta_{-} = \frac{1}{b_{\textit{gen}}}, 
\quad
2 \beta_0 = \beta_{+} + \beta_{-}
\end{equation}

\subsection{AGT parameterisation. Minimal models}
\label{parameterisation.minimal.models}

A minimal model $\cM^{\, p, \pp}$, based on a Virasoro algebra $\cV^{p, \pp}$, characterised 
by a central charge $c_{p, \pp} < 1$, that we parameterise as

\begin{equation}
\label{central.charge.minimal}
c_{p, \pp} = 1 - 6 \ll a_{p, \pp} - \frac{1}{a_{p, \pp}} \rr^2,
\quad
a_{p, \pp} = \ll \frac{\pp}{p} \rr^{\frac{1}{2}},
\end{equation}

\noindent where $\{p, \pp\}$ are the minimal model parameters, which are co-prime integers and 
satisfy $0 < p < \pp$, in our conventions. In the Coulomb gas approach to computing conformal blocks 
in minimal models with $c < 1$ \cite{nienhuis, dotsenko.fateev}, the screening charges 
$\{\apos, \aneg\}$, and the background charge, $- 2 \alpha_0$, satisfy 

\begin{equation}
\label{minimal.model.charges}
\apos =            a_{p, \pp}, 
\quad
\aneg = - \ \frac{1}{a_{p, \pp}},
\quad
2 \alpha_0 = \apos + \aneg
\end{equation}

\noindent The AGT parameterisation of $\cM^{\, p, \pp, \cH}$ is obtained by choosing 

\begin{equation}
\label{neg.0.pos}
\epsilon_1 < 0 < \epsilon_2, 
\quad
\epsilon_1 = \aneg, 
\quad
\epsilon_2 = \apos 
\end{equation}

\noindent so that $\aneg < 0 < \apos$. Since we focus on $\cM^{\, p, \pp, \cH}$, 
we work in terms of $\{\aneg, \apos\}$ instead of $\{ \epsilon_1, \epsilon_2\}$, 
and write the elementary function $E[x, Y_i, W_j, \square]$ as 

\begin{equation}
\label{E.minimal}
\boxed{
E[x, Y_i, W_j, \square] =
x
+ A^{+}_{\square, i}  \, \apos
- L^{ }_{\square, j}  \, \aneg
}
\end{equation}

\subsection{Charge content}
\label{charge.content}
We need two distinct objects that, in Coulomb gas terms, are expressed in terms of the 
screening charges $\{\apos, \aneg\}$. 
{\bf 1.} The charge $\mu_{r, s}$ of the vertex operator ${\cO}_{\mu}$ that 
intertwines two irrep's $\cH^{p, \pp}_{r_1, s_1}$ 
and $\cH^{p, \pp}_{r_2, s_2}$, and 
{\bf 2.} The highest weight $|a_{r, s} \rangle$ of an irrep $\cH^{p, \pp}_{r, s}$. 
Following \cite{alday.gaiotto.tachikawa, alba.fateev.litvinov.tarnopolskiy}, we use 
$\{r, s\}$ as indices for the charge $\mu_{r, s}$ of the vertex operator $\cO_{\mu_{r, s}}$, and 
$\{r, s\}$ as indices for the charge $a_{r, s}$      of the highest weight $| a_{r, s} \rangle$. 
These charges are parameterised in terms of $\apos$ and $\aneg$ as follows 

\begin{multline}
\label{parameters.03}
\mu_{r, s} =   - \ll \frac{r-1}{2} \rr \, \apos - \ll \frac{s-1}{2} \rr \, \aneg, \quad 
\\
     a_{r, s} = - \frac{r  }{2} \ \apos - \frac{s}{2}   \ \aneg, \quad 
1 \leq r \leq   p - 1,
1 \leq s \leq \pp - 1
\end{multline}

\noindent Note that the same numerical values of $\{r, s\}$ indicate different charge contents in
$\mu_{r, s}$ and in $a_{r, s}$. In particular,

\begin{equation}
\label{relations.01}
2 \mu_{r, s} = 2 a_{r, s} + \ll \apos + \, \aneg \rr
\end{equation}

\subsection{Unrestricted instanton partition functions give incorrect 1-point functions on the torus} 
\label{unmodified.torus}
Consider a conformal block in a 1-point function in $\cM^{\, p, \pp, \cH}$ on a torus, Figure {\bf 4}.

%
\begin{center}
\begin{minipage}{5.0in}
\setlength{\unitlength}{0.001cm}
\renewcommand{\dashlinestretch}{30}
\begin{picture}(4800, 3400)(-4000, 0500)
%
\thicklines
%
\put(2200,2200){\circle{2200}}
\path(2200,3300)(2200,3900)
 \put(2400,3600){$\mu$}
\put(3500,2200){$\cH^{p, \pp}_{r, s} \otimes \cF$}
\end{picture}
\begin{ca}
\label{1.point.function.on.torus}
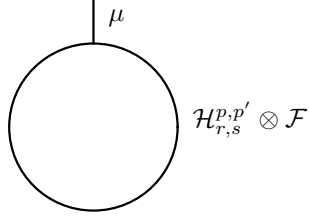
A graph of a conformal block of a 1-point function on a torus. The vertex operator 
insertion carries a charge $\mu$. The chiral states that flow in the channel belong 
to $\cH^{p, \pp}_{r, s} \otimes \cF$. For this 1-point function to be finite, 
the fusion rules must be satisfied.
\end{ca}
\end{minipage}
\end{center}
\bigskip

Following \cite{alday.gaiotto.tachikawa}, this is given by the instanton partition 
function of the ${\mathcal N} = 2^{\star}$ $U(2)$ theory, 

\begin{equation}
\label{torus.01}
Z_{Nek}^{{\mathcal N} = 2^{\star}, U(2)} \ll \vec a, \vec \mu \rr = 
\sum_{\vec Y} q^{|{\vec Y}|} 
\frac{
z_{num} \ll \vec a, \vec Y \ | \ \mu \ | \ \vec a, \vec Y \rr
}
{
z_{den} \ll \vec a, \vec Y \ |           \ \vec a, \vec Y \rr
},
\end{equation}

\noindent where $\mu$ is determined by the operator insertion, and ${\vec a}$ is determined 
by the states of the $\cH^{p, \pp}_{r, s}$ that flow in the torus 
and determine the conformal block. When the inserted operator is the identity, that is 
$\{r, s\}$, then 
\footnote{\ See subsection {\bf \ref{notation.02}}.} 
$\mu = 0$, and if ${\vec Y}$ is an unrestricted partition pair as in the original AGT prescription, 
then

\begin{equation}
\label{torus.02}
Z_{Nek}^{{\mathcal N} 2^{\star}, U(2)} \ll \vec a, \vec 0 \rr
= \sum_{\vec Y} q^{|{\vec Y}|}  
= \frac{1}{\prod_{n=1}^{\infty} \ll 1 - q^n \rr^2}
\end{equation}

\noindent while the correct result is

\begin{equation}
\label{torus.03}
Z_{Nek}^{{\mathcal N} 2^{\star}, U(2)} \ll \vec a, \vec 0 \rr = 
\frac{ 
\chi^{p, \pp}_{r, s} 
}{
\prod_{n=1}^{\infty} \ll 1 - q^n\rr
},
\end{equation}

\noindent where $\chi^{p, \pp}_{r, s}$ is the character of 
the irrep $\cH^{p, \pp}_{r, s}$ that flows in the torus, 
and $\vec a = \{a, -a\}$, 
where $2a = - r \, \apos - s \, \aneg$,

\begin{equation}
\label{virasoro.character.01}
\chi^{p, \pp}_{r, s} = 
\frac{
\sum_{k = - \infty}^{\infty}
\ll q^{k^2 p \pp + k (\pp r - p s)} - q^{(k p + r) (k \pp + s)} \rr
}
{
\prod_{n=1}^{\infty} \ll 1 - q^n \rr
}
\end{equation}

This simple example makes it clear that applying the prescription of 
\cite{alday.gaiotto.tachikawa} to $\cM^{\, p, \pp, \cH}$ without 
modification, leads to incorrect answers. In the following section, 
we find that it leads to zeros in the denominators of the summands.

\section{Restricted instanton partition functions for $\cM^{\, p, \pp, \cH}$. 
\newline 
The denominator}
\label{modified.denominator}

Consider the denominator $z_{den}$ of $Z_{bb}$ in (\ref{z.bb}). To look for zeros 
in $z_{den}$, it is sufficient to look for zeros in $z_{norm} [\vec a, \vec Y]$ 
in (\ref{z.norm}). Consider $\cB^{\, p, \pp, \cH}_n$ and focus on a channel that carries 
states that belong to $\cH^{p, \pp}_{r, s}$. 

\begin{Proposition} 
\label{proposition.01}
$z_{norm} [\vec a, \vec Y] \neq 0$, if and only if 
\begin{equation}
\label{conditions.01.simpler}
Y_{2, \rrho} - Y_{1, \rrho +    s      -1} \geq 1 -     r, 
\quad
Y_{1, \rrho} - Y_{2, \rrho + [\pp - s] -1} \geq 1 - \ll p - r \rr, 
\end{equation}
\noindent where $Y_{i, \rrho}$ is row-$\rrho$ in $Y_i$, $i \in \{1, 2\}$. 
\end{Proposition}

The proof of Proposition {\bf \ref{proposition.01}} is based on checking the products 
that appear in $z_{norm} [\vec a, \vec Y]$ for zeros. 

\subsection{More notation} 
We set $a_1 = - a_2 = a$, and $a_1 - a_2 = 2a$. If a channel $\chi_{\i}$ carries states from
$\cH^{p, \pp}_{r_{\i}, s_{\i}} \otimes \cF$, then the label $a_{\i}$ of the corresponding highest 
weight is

\begin{equation}
\label{highest.weight.in.channel}
a_{r_{\i}, s_{\i}} = - r_{\i} \, \apos - s_{\i} \, \aneg 
\end{equation}

\subsection{Two zero-conditions}
\label{to.vanish}
In the sequel, we find that an instanton partition function has a zero when an equation 
of type

\begin{equation}
\label{zeros.example.01}
C_{+} \, \apos + C_{-} \, \aneg = 0,
\end{equation}

\noindent where $\aneg < 0 < \apos$, is satisfied. Equivalently, an instanton partition function  
has a zero when the two conditions

\begin{equation}
\label{zeros.example.02}
C_+ =  c   \ p, 
\quad
C_- =  c \ \pp, 
\end{equation}

\noindent are satisfied, where $c$ is some constant that needs to be determined. 
Given two conditions, such as (\ref{zeros.example.02}), we need, for the purposes 
of comparing with known results, to re-write them as one condition. 

\subsection{From two zero-conditions to one zero-condition} 
\label{from.2.to.1}

Consider the two conditions

\begin{equation}
\label{conditions.example.01}
\boxed{
  A^{ }_{\square, 1}   =  A^{\prime} \geq 0,
\quad 
- L^{ }_{\square, 2}   =  L^{\prime} \geq 0
}
\end{equation}

\noindent which are satisfied if $\square \in Y_1$, and $\square \not \in Y_2$
\footnote{\ We chose the labels of the Young diagrams to be concrete. The same arguments 
apply under $Y_1\Leftrightarrow Y_2$.}. 
If $\square$ is in row-$\rrho$ and column-$\ssigma$ in $Y_1$, then the first condition in 
(\ref{conditions.example.01}) implies that there is a cell $\boxplus \in Y_1$, to the right of 
$\square$, with coordinates $\{\rrho, \ssigma + A^{\prime}\}$, that lies on a vertical boundary. 
In other words, {\bf 1.} there are no cells to the right of $\boxplus$, and {\bf 2.} there may 
or may not be cells below $\boxplus$. 
This means that 
column-$(\ssigma + A^{\prime})$ in $Y_1$, or equivalently, 
row-$(\ssigma + A^{\prime})$    in $Y_1^{\intercal}$, 
has length {\it at least} $\rrho$,
\footnote{\ All equations and inequalities in the sequel involve rows of Young diagrams, 
and never columns. For example, $Y^{\intercal}_{1, \ssigma + A^{\prime}}$ is 
row-$(\ssigma + A^{\prime})$ in diagram $Y^{\intercal}_1$, which is the transpose of $Y_1$. 
The subscript $\ssigma$ is there only because the corresponding row is a column in a diagram 
that we started our arguments with.
}, 

\begin{equation}
\label{conditions.example.02}
Y^{\intercal}_{1, \ssigma + A^{\prime}} \geq \rrho
\end{equation}

\noindent From the definition of $L_{\square, 2}$, we write the second condition in 
(\ref{conditions.example.01}) as 
$- L_{\square, 2}   = L^{\prime} = \rrho - Y^{\intercal}_{2, \ssigma}$,
that is 
$\rrho = L^{\prime} + Y^{\intercal}_{2, \ssigma}$, 
and using (\ref{conditions.example.02}), we obtain 
$Y^{\intercal}_{1, \ssigma + A^{\prime}} \geq L^{\prime} + Y^{\intercal}_{2, \ssigma}$,
which we choose to write as

\begin{equation}
\label{conditions.example.04}
Y^{\intercal}_{1, \ssigma + A^{\prime}} - Y^{\intercal}_{2, \ssigma} \geq L^{\prime},
\end{equation}

\noindent which is one condition that is equivalent to the two conditions in 
(\ref{conditions.example.01}). 

\subsection{One non-zero condition}
\label{1.not.to.vanish}
Consider a function $z[Y_1, Y_2]$, of a pair of Young diagrams $\{Y_1, Y_2\}$, such that $z[Y_1, Y_2] = 0$, 
{\it if and only if} (\ref{conditions.example.04}) is satisfied. This implies that $z[Y_1, Y_2] \neq 0$, 
{\it if and only if} $\{Y_1, Y_2\}$ satisfies the complementary condition

\begin{equation}
\label{conditions.example.05}
Y^{\intercal}_{1, \ssigma + A^{\prime}} - Y^{\intercal}_{2, \ssigma} < L^{\prime},
\end{equation}

\noindent which we choose to write as 

\begin{equation}
\label{conditions.example.06}
\boxed{
Y^{\intercal}_{2, \ssigma} - Y^{\intercal}_{1, \ssigma + A^{\prime}} \geq 1 - L^{\prime}
}
\end{equation}

\subsubsection{Remark} 
Since we use equations such as (\ref{conditions.example.01}) and (\ref{conditions.example.06}) 
frequently in the sequel, refer 
to the former as {\it \lq zero-conditions\rq}, and 
to the latter as {\it \lq non-zero-conditions\rq}. 

\subsection{Products that appear in the denominator} 
\label{products.denominator}
Two types of products appear in $z_{norm}$, 
{\bf 1.} products in the form 
$\prod_{\square      \in Y_i}                  E[a_i - a_j, Y_i, Y_j, \square]$ 
that we refer to as  $\{Y_i, Y_j\}_{\it den}$, 
and
{\bf 2.} products in the form 
$\prod_{\square      \in Y_i} [\apos + \aneg - E[a_i - a_j, Y_i, Y_j, \square]]$
that we refer to as $\{Y_i, Y_j\}_{\it den}^{\prime}$. 

\subsection{In search of zeros}
\label{looking.for.zero}
In the following subsections, 
{\bf 1.} we consider the products that appear in $z_{den}$, one at a time, 
{\bf 2.} we search for possible zeros, as in subsection {\bf \ref{to.vanish}}, 
{\bf 3.} we find the conditions that we need to impose on the pair $\{Y_1, Y_2\}$ 
in order to avoid the zeros, and 
{\bf 4.} when there is more than one set of conditions to avoid the zeros, we choose 
the stronger set. That is, the set that ensures that all zeros are eliminated. We use 
the fact that $r$, $s$, $p - a$ and $\pp - s$ are non-zero positive integers.

\subsection{$ \mathbf{\{Y_1, Y_1\}_{\it den}}$} 
\label{den.11}
This product does not vanish, since this requires that there is a factor that 
satisfies

\begin{equation}
\label{zeros.11}
E[0, Y_1, Y_1, \square] 
=
  A^{+}_{\square, 1} \  \apos
- L^{ }_{\square, 1} \  \aneg
=
0,
\end{equation}

\noindent which is not possible since $\square \in Y_1$ and $\aneg < 0$.

\subsection{$ \mathbf{\{Y_1, Y_1\}^{\prime}}$, $ \mathbf{\{Y_2, Y_2\}_{\it den}} $ and $ \mathbf{\{Y_2, Y_2\}^{\prime}_{\it den}} $} 
\label{den.conjugate.11.den.22.and.den.conjugate.22}
These products do not vanish for the same reason that $\{Y_1, Y_1\}_{\it den}$ 
in paragraph {\bf \ref{den.11}} does not vanish. 

\subsection{ $\mathbf{\{Y_1, Y_2\}_{\it den}} $}
\label{den.12}
This product vanishes if any factor satisfies 
\begin{equation}
\label{zeros.12}
E[- r \, \apos - s \, \aneg, Y_1, Y_2, \square]
=
\\
\ll  - r + A^{+}_{\square, 1} \rr \, \apos
+
\ll  - s - L^{ }_{\square, 2} \rr \, \aneg
=
0,
\end{equation}
\noindent which lead to the conditions
\begin{equation}
\label{two.zero.conditions.12}
  A^{ }_{\square, 1} = r - 1 + c\  p 
\quad 
- L^{ }_{\square, 2} = s     + c\ \pp 
\end{equation}

\noindent Since $A_{\square, i}$, $L_{\square, i}$, $i \in \{1, 2\}$, $r$, and $s$
are non-zero positive integers, and $p$ and $\pp$ are positive co-primes, $c$ must 
be a non-negative integer, and the conditions in (\ref{two.zero.conditions.12}) are 
possible for $c = \{0, 1, \cdots\}$, $\square \in Y_1$ and $\square \not \in Y_2$
\footnote{\ Note that from conditions (\ref{two.zero.conditions.12}), if 
the Young diagram $Y$ such that $\square      \in Y$, which in this case is $Y_1$, 
is sufficiently large compared to 
the Young diagram $W$ such that $\square \not \in W$, which in this case is $Y_2$, 
then the product under discussion will have more than one zero. This will be the 
case in the rest of the factors discussed in this section as well.}.  

\subsubsection{From two zero-conditions to one non-zero-condition} 
\label{translating.01}
Following paragraphs {\bf \ref{from.2.to.1}} and {\bf \ref{1.not.to.vanish}}, the two zero-conditions 
in (\ref{two.zero.conditions.12}) can be translated to one non-zero-condition, 

\begin{equation}
\label{one.non.zero.condition.12}
Y^{\intercal}_{1, \ssigma} - Y^{\intercal}_{2, \ssigma + r - 1 + c\, p} \geq 1 - s - c\, \pp 
\end{equation}

\subsubsection{The stronger condition} 
\label{the.stronger.condition.01}
Equation (\ref{one.non.zero.condition.12}) is the statement that to eliminate the zeros, we want 
$Y^{\intercal}_{2, \ssigma} - Y^{\intercal}_{1, \ssigma + r - 1 + c\, p} \geq 1 - s - c\, \pp$, 
where $c = \{0, 1, \cdots\}$
Since the row-lengths of a partition are by definition weakly decreasing, and 
$c = \{0, 1, \cdots\}$, this is the case if 
$Y^{\intercal}_{2, \ssigma} - Y^{\intercal}_{1, \ssigma + r - 1}         \geq 1 - s - c\, \pp$, 
which is the case if 
$Y^{\intercal}_{2, \ssigma} - Y^{\intercal}_{1, \ssigma + r - 1}         \geq 1 - s$. 
Thus, we should set $c=0$, and obtain 

\begin{equation}
\label{Burge.condition.01}
Y^{\intercal}_{2, \ssigma} - Y^{\intercal}_{1, \ssigma + r - 1} \geq 1 - s 
\end{equation}

\subsection{The 1st Burge condition}
\label{first.Burge.condition.subsection}
Condition (\ref{Burge.condition.01}) says that if we delete the top $(r-1)$ rows 
and the left-most $(s-1)$ columns of $Y^{\intercal}_1$, to obtain a reduced Young 
diagram that we refer to as $Y^{\intercal}_{red, 1}$, then 

\begin{equation}
\label{reduction.01}
Y^{\intercal}_{2, \ssigma} - Y^{\intercal}_{red, 1, \ssigma} \geq 0 
\end{equation}

\noindent which implies

\begin{equation}
\label{reduction.02}
Y_{2, \rrho} - Y_{red, 1, \rrho} \geq 0
\end{equation}

\noindent where $Y_{\rrho}$ is row-$\rrho$ of $Y$. Transposing the $(r-1)$ rows 
and $(s-1)$ columns that we removed earlier from 
$Y^{\intercal}_1$ in order to obtain $Y^{\intercal}_{red, 1}$, we obtain 
$(r-1)$ columns and $(s-1)$ rows that we can add to the top and to the left of 
$Y_{1, red}$, respectively, to obtain

\begin{equation}
\label{first.Burge.condition}
\boxed{
Y_{2, \rrho} - Y_{1, \rrho + s - 1} \geq 1 - r
}
\end{equation}

\noindent which is the 1st Burge condition in (\ref{Burge.conditions}).

\subsection{ $\mathbf{\{Y_2, Y_1\}_{\it den}} $}
\label{den.21}
This product vanishes if any factor satisfies 
\begin{equation}
\label{zeros.21}
E[ r \, \apos + s \, \aneg, Y_2, Y_1, \square]
=
\ll r + A^{+}_{\square, 2} \rr \, \apos 
+ 
\ll s - L^{ }_{\square, 1} \rr \, \aneg
=
0,
\end{equation}
\noindent which leads to the conditions
\begin{equation}
\label{conditions.21.01}
  A^{ }_{\square, 2} = - 1 - r + c\   p, 
\quad
- L^{ }_{\square, 1} =     - s + c\ \pp,
\end{equation}
\noindent which are possible for $c = \{1, 2, \cdots\}$, $\square \in Y_2$ and $\square \not \in Y_1$.

\subsubsection{From two zero-conditions to one non-zero-condition} 
\label{translating.02}
Following paragraphs {\bf \ref{from.2.to.1}} and {\bf \ref{1.not.to.vanish}}, the two zero-conditions 
in (\ref{conditions.21.01}) can be translated to one non-zero-condition, 
\begin{equation}
\label{condition.21.04}
Y^{\intercal}_{1, \ssigma} - Y^{\intercal}_{2, \ssigma - 1 - r + c\, p} \geq 1 + s - c\, \pp 
\end{equation}
\subsubsection{The stronger condition} 
\label{the.stronger.condition.02}
Equation (\ref{condition.21.04}) is the statement that to eliminate the zeros, 
we want 
$Y^{\intercal}_{2, \ssigma} - Y^{\intercal}_{1, \ssigma - 1 - r + c\, p} \geq 1 + s - c\, \pp$, 
where $c = \{1, 2, \cdots\}$.
Since the row-lengths of a partition are by definition weakly decreasing, and 
$c = \{1, 2, \cdots\}$, this is the case if 
$Y^{\intercal}_{2, \ssigma} - Y^{\intercal}_{1, \ssigma + p - r - 1}         \geq 1 - s - c\, \pp$, 
which is the case if 
$Y^{\intercal}_{2, \ssigma} - Y^{\intercal}_{1, \ssigma + p - r - 1}         \geq 1 - p + s$. 
Thus, we should set $c=1$, to obtain 
\begin{equation}
\label{Burge.condition.02}
Y^{\intercal}_{1, \ssigma} - Y^{\intercal}_{2, \ssigma + [p - r] - 1} \geq 1 - \ll \pp - s \rr
\end{equation}

\subsection{The 2nd Burge condition} 
\label{second.Burge.condition.subsection}
Following the argument in subsection {\bf \ref{first.Burge.condition.subsection}}, we can write 
condition (\ref{Burge.condition.02}) as

\begin{equation}
\label{second.Burge.condition}
\boxed{
Y_{1, \rrho} - Y_{2, \rrho + [\pp - s] - 1} \geq 1 - \ll p - r \rr
}
\end{equation}

\noindent which is the 2nd Burge condition in (\ref{Burge.conditions}).


\subsection{$ \mathbf{\{Y_1, Y_2\}_{\it den}^{\prime}} $}
\label{den.conjugate.12}
This product vanishes if any factor satisfies

\begin{multline}
\label{zeros.conjugate.12}
- \, \apos - \, \aneg + E[- r \, \apos - s \, \aneg, Y_1, Y_2, \square] 
=
\\
\ll - r + A^{ }_{\square, 1} \rr \, \apos 
+
\ll - s - L^{+}_{\square, 2} \rr \, \aneg
= 0,
\end{multline}

\noindent which leads to the conditions

\begin{equation}
\label{conditions.conjugate.12.01}
  A_{\square, 1}   =     r + c\, p,  
\quad
- L_{\square, 2}   = 1 + s + c\,\pp,
\end{equation}

\noindent which, using the same arguments as in subsections {\bf \ref{den.12}} and {\bf \ref{den.21}}, 
are possible for $c = \{0, 1, \cdots\}$, $\square \in Y_1$, $\square \not \in Y_2$, and we should
choose $c=0$ to obtain  

\begin{equation}
\label{weaker.than.Burge.condition.01}
Y^{\intercal}_{2, \ssigma} - Y^{\intercal}_{1, \ssigma + r} \geq - s 
\end{equation}

\noindent Comparing condition (\ref{Burge.condition.01}) with condition (\ref{weaker.than.Burge.condition.01}), 
we see that the former is stronger than the latter, for the same reasons 
as in paragraph {\bf \ref{the.stronger.condition.01}}. Thus this case does not offer new conditions 
on the partition pair. 

\subsection{$ \mathbf{\{Y_2, Y_1\}_{\it den}^{\prime}} $}
\label{den.conjugate.21}
This product vanishes if any factor satisfies  

\begin{multline}
\label{zeros.conjugate.21}
- \, \apos - \, \aneg + E[r \, \apos + s \, \aneg, Y_2, Y_1, \square] =
\\
  \ll   r + A^{ }_{\square, 2} \rr \   \apos 
+ \ll   s - L^{+}_{\square, 1} \rr \   \aneg  = 0,
\end{multline}

\noindent which leads to the conditions

\begin{equation}
\label{conditions.conjugate.21.01}
  A^{ }_{\square, 2} =   - r + c\  p, 
\quad
- L^{ }_{\square, 1} = 1 - s + c\ \pp, 
\end{equation}

\noindent which, using the same arguments as in subsections {\bf \ref{den.12}} and {\bf \ref{den.21}}, are 
possible for $c = \{1, 2, \cdots\}$, $\square \in Y_2$ and $\square \not \in Y_1$, and we should choose $c=1$
to obtain

\begin{equation}
\label{weaker.than.Burge.condition.02}
Y^{\intercal}_{2, \ssigma} - Y^{\intercal}_{1, \ssigma + [p - r]} \geq - \ll \pp - s \rr
\end{equation}

\noindent Comparing condition (\ref{Burge.condition.02}) with condition (\ref{weaker.than.Burge.condition.02}),
we see that the former is stronger than the latter, for the same reasons
as in paragraph {\bf \ref{the.stronger.condition.02}}. Thus this case does not offer new conditions
on the partition pair.

\subsection{Restricted instanton partition functions give the correct 1-point function on the torus} 
\label{modified.torus}
From the discussion in paragraphs {\bf \ref{den.11}--\ref{den.conjugate.21}},
we conclude that $z_{den}$ has no zeros if the conditions in 
(\ref{Burge.condition.01}) and 
(\ref{Burge.condition.02}) are satisfied.
As mentioned in section {\bf \ref{introduction}}, these conditions on partition pairs are known. They 
were introduced and studied in \cite{burge}, and were further studied and called {\it Burge pairs} in 
\cite{foda.lee.welsh}. A full and explicit derivation of the fact that the generating function of 
the Burge pairs, that satisfy conditions 
(\ref{Burge.condition.01}) and
(\ref{Burge.condition.02}), is the $q$-series in (\ref{torus.03}), we refer the reader to 
Appendix {\bf A} of \cite{foda.lee.welsh}. 

\subsubsection{Remark} The conditions obtained in this note were written 
differently in \cite{burge, foda.lee.welsh} for three reasons. 
{\bf 1.} These papers used the notation 
$\{a, b, \alpha, \beta   \}$, which in terms of the variables 
$\{r, s,  p,     \pp     \}$ used in this work are 
$a=r, b=p-r, \alpha = s$, and $\beta = \pp - s$,
{\bf 2.} The partition rows were labeled such that $Y_i \geq Y_{i-1}$, 
while in this note, we assume the opposite (and more conventional) labeling, and 
{\bf 3.} The conventions in \cite{burge, foda.lee.welsh} are such that the conditions were expressed 
in terms of the Young diagrams that the presentation naturally started with, while in this work, we 
wished to follow the conventions of 
\cite{alday.gaiotto.tachikawa, nekrasov, alba.fateev.litvinov.tarnopolskiy}, so we ended up expressing 
the conditions on the partition pair $\{Y_1, Y_2\}$ as conditions on $\{Y^{\intercal}_1, Y^{\intercal}_2\}$.

\section{Restricted instanton partition functions for $\cM^{\, p, p^{\prime}, \cH}$. 
\newline 
The numerator}
\label{modified.numerator}

\subsection{Products that appear in the numerator}
\label{classification.of.products.denominator}
Two types of products appear in $z_{num}$,
{\bf 1.} products in the form
$\prod_{\square \in Y_i} [                 E[ x, Y_i, W_j, \square] - \mu]$
that we refer to as  $\{Y_i, W_j\}_{num}$,
and
{\bf 2.} products in the form
$\prod_{\square  \in W_j} [\apos + \aneg - E[-x, W_j, Y_i, \square] - \mu]$
that we refer to as $\{W_j, Y_i\}_{num}^{\prime}$.
We need to examine the conditions that each of these factors imposes on the
partition pairs $\{Y_1, Y_2\}$ and $\{W_1, W_2\}$. 

\subsection{Notation} 
\label{notation.02}
We set

\begin{multline}
\label{notation.02.01}
a_1 =  - a_2 = a = - \ll \frac{m_a + 1}{2} \rr    \ \apos - \ll \frac{n_a + 1}{2} \rr   \ \aneg, 
\\
m_a \in \{0, 1, \cdots,     p - 2\}, 
n_a \in \{0, 1, \cdots,   \pp - 2\}, 
\end{multline}

\begin{multline}
\label{notation.02.02}
b_1 =  - b_2 = b = - \ll \frac{m_b + 1}{2} \rr    \ \apos - \ll \frac{n_b + 1}{2} \rr   \ \aneg, 
\\
m_b \in \{0, 1, \cdots,   p - 2\},
n_b \in \{0, 1, \cdots, \pp - 2\},
\end{multline}

\begin{equation}
\label{notation.02.03}
\mu  =              - \frac{1}{2} m_{\mu} \ \apos - \frac{1}{2} n_{\mu} \ \aneg,
\quad
m_{\mu} \in \{0, 1, \cdots,   p-2\}, 
n_{\mu} \in \{0, 1, \cdots, \pp-2\}, 
\end{equation}

\noindent and use the subscript $a$ $(b)$ to indicate the parameters that appear in the conditions 
on the partition pairs $\vec Y$ $(\vec W)$ that label the states in the incoming (outgoing) channel 
that flows towards (away from) the vertex operator insertion. It is useful to note that, in this 
notation,

\begin{equation}
r_a = m_a + 1, \quad 
s_a = n_a + 1, \quad 
r_b = m_a + 1, \quad 
s_b = n_a + 1
\end{equation}

Further, to simplify the presentation, we use the notation

\begin{equation}
\label{M.N}
M_{[\pm, \pm, \pm]} = \frac{1}{2} \ll \pm m_a \pm m_b \pm m_{\mu} \rr, 
\quad
N_{[\pm, \pm, \pm]} = \frac{1}{2} \ll \pm n_a \pm n_b \pm n_{\mu} \rr
\end{equation}

\subsection{The fusion rules}

In the notation of subsection {\bf \ref{notation.02}}, the fusion rules are

\begin{equation}
\label{fusion.rules}
m_a + m_b + m_{\mu} = 0  \ \textit{mod}  \ 2, \quad
n_a + n_b + n_{\mu} = 0  \ \textit{mod}  \ 2
\end{equation}

\noindent and the triple $\{m_a, m_b, m_{\mu}\}$ satisfies the triangular conditions

\begin{equation}
\label{triangular.conditions}
m_a + m_b     \geq m_{\mu}, \quad 
m_b + m_{\mu} \geq m_a,     \quad 
m_{\mu} + m_a \geq m_b
\end{equation}

\noindent with analogous triangular conditions for $\{n_a, n_b, n_{\mu}\}$. 

\subsection{Bounds on $M_{[\pm, \pm, \pm]}$ and $N_{[\pm, \pm, \pm]}$}
For the purposes of the proofs in subsections {\bf \ref{remaining.six.products}} and 
{\bf \ref{if.den.is.zero.then.num.is.zero}}, we need to show that 
$M_{[\pm, \pm, \pm]}$ and 
$N_{[\pm, \pm, \pm]}$ satisfy the bounds

\begin{equation}
\label{bounds.on.M.and.N}
0 \leq M_{[\pm, \pm, \pm]} \leq   p - 2,
\quad
0 \leq N_{[\pm, \pm, \pm]} \leq \pp - 2,
\end{equation}

\noindent The lower bounds follow from the lower bounds in the definitions 
(\ref{notation.02.01}--\ref{notation.02.03}). The upper bounds are obtained as follows.
There are two ways to choose the charge content of the highest weight state of a Virasoro 
irrep. The first choice is $\alpha_{r, s}$, where 

\begin{multline}
2 \alpha_{r, s} = - \ll r - 1 \rr \apos - \ll s - 1 \rr \aneg, 
\\
r = m + 1, \quad s = n + 1,
\\
0 \leq m \leq p - 2, \quad 0 \leq n \leq \pp - 2,
\end{multline}

\noindent while the second choice is

\begin{multline}
2 \alpha_{r^{\prime}, s^{\prime}} = - \ll r^{\prime} - 1 \rr \apos - \ll s^{\prime} - 1 \rr \aneg,
\\
r^{\prime} = m^{\prime} + 1, \quad s^{\prime} = n^{\prime} + 1,
\\
0 \leq m^{\prime} \leq p - 2, \quad 0 \leq n^{\prime} \leq \pp - 2,
\end{multline}

\noindent and the two choice are related by

\begin{equation}
\label{r.prime.s.prime}
r^{\prime} =   p - r, 
\quad
s^{\prime} = \pp - s 
\end{equation}

\noindent While the two representations are the same, for the purposes of the proofs in the 
sequel, we need to use one or the other, as follows.

Scanning a linear conformal block $\cB^{p, \pp, \cH}_n$ from left to right, one considers 
the building block $Z_{bb}^{\i}$, $\i = 1, 2, \cdots$, with the Virasoro irrep labeled by
$\{r_{\i - 1}, s_{\i - 1}\}$ flowing in from the left, 
the vertex operator $\cO_{\i}$ of the primary field labeled by 
$\{r_{\mu}, s_{\mu}\}$ in the middle, and the Virasoro irrep labeled by
$\{r_{\i}, s_{\i}\}$ flowing out to the right.
Suppose that the charge content of the incoming primary field in $\chi_{\i - 1}$ is fixed
\footnote{\ Starting from $Z_{bb}^1$, we can choose the charge of the highest weight state
in $\cO_0$ either way, but for the purposes of this proof, it is sufficient to consider 
an arbitrary $Z_{bb}^{\i}$, $\i = 1, 2, \cdots, n+1$, and take the charge of the primary 
field in $\chi_{\i - 1}$ to be fixed.}. 
The charge content of primary state $\mu$ of the vertex operator $\cO_{\mu}$ in the middle,
and that of the outgoing primary field in $\chi_{\i}$ are not fixed yet, and 
each can be chosen in one of two equivalent ways. We wish to show that we can choose these 
charge contents in such a way that that the upper bounds in (\ref{bounds.on.M.and.N}) are 
satisfied. This will simplify our proofs in the sequel.

If $\{m_{\i}, n_{\i}\}$ and $\{m_{\mu}, n_{\mu}\}$ are such that the upper bounds
in (\ref{bounds.on.M.and.N}) are satisfied, then use this choice. 
If $\{m_{\i}, n_{\i}\}$ and $\{m_{\mu}, n_{\mu}\}$ are such that the upper bounds
in (\ref{bounds.on.M.and.N}) are {\it not} satisfied, we choose the dual representation 
of the vertex operator in the middle and the outgoing Virasoro irrep
\footnote{\ Remember that the charge content of the incoming primary field is given 
and cannot be changed.}. 
In other words,
$p - \frac{1}{2} \ll m_{\i - 1} + m_{\mu}          + m_{\i}          \rr - 2$,
that does not satisfy the upper bound, becomes

\begin{multline}
\label{upper.bound.on.M}
p - \frac{1}{2} \ll m_{\i - 1} + m_{\mu}^{\prime} + m_{\i}^{\prime} \rr - 2 =
\\
p - \frac{1}{2} \ll m_{\i - 1} + p - m_{\mu} - 2  + p - m_{\i} - 2  \rr - 2 =
\\
    \frac{1}{2} \ll m_{\i - 1} + m_{\mu} + m_{\i} \rr \geq 0
\end{multline}

\noindent using the triangular conditions (\ref{triangular.conditions}), and similarly 

\begin{equation}
\label{upper.bound.on.N}
\pp - \frac{1}{2} \ll n_{\i - 1} + n_{\mu}^{\prime} + n_{\i}^{\prime} \rr - 2 \geq 0
\end{equation}

\noindent Now the charge content of the outgoing primary field is fixed and goes on to become 
the incoming primary field of $Z_{bb}^{\i + 1}$ or the primary state of $\cO_{n + 3}$.
Thus we can always choose the charge contents such that the upper bounds in equations 
(\ref{upper.bound.on.M}) and (\ref{upper.bound.on.N}).

In the following subsections, we consider the conditions that products in the numerator 
must satisfy to be non-zero 
\footnote{\ In writing equations 
(\ref{11.num.1.condition}),
(\ref{11.num.prime.1.condition}), and
(\ref{22.num.1.condition}--\ref{21.num.1.condition}), 
we choose to group terms together in such a way to make the analogy 
with conditions 
(\ref{Burge.condition.01}--\ref{weaker.than.Burge.condition.02}),
that involve $\{Y_1, Y_2\}$ only, relatively more clear. Basically, 
$M$ and $N$ in the former are analogues of $(r-1)$ and $(s-1)$ in 
the latter.}.

\subsection{$\mathbf{\{Y_1, W_1\}_{num}}$}
\label{11.num}
This product vanishes if any factor satisfies

\begin{multline}
E[ a - b, Y_1, W_1, \square] - \mu = 
\\
- M_{[+, -, -]} \, \apos
- N_{[+, -, -]} \, \aneg
+ A^{+}_{Y_1}   \, \apos - L^{ }_{W_1}   \, \aneg
= 0,
\end{multline}

\noindent which leads to the zero-conditions

\begin{equation}
\label{11.num.2.conditions}
  A_{\square, Y_1}   = - M_{[-, +, +]} + c\  p - 1,
\quad
- L_{\square, W_1}   = - N_{[-, +, +]} + c\ \pp 
\end{equation}

\noindent From the triangular conditions (\ref{triangular.conditions}), 
the maximal value of 
$M_{[-, +, +]}$ is   $p - 2$, and the maximal value of 
$N_{[-, +, +]}$ is $\pp - 2$, 
thus the stronger condition corresponds to $c=1$, and we obtain two 
zero-conditions that we can write as one non-zero-condition,

\begin{equation}
\label{11.num.1.condition}
W^{\intercal}_{1, \ssigma} - Y^{\intercal}_{1, \ssigma + [p - M_{[-, +, +]} - 1]} \geq - \ll \pp - N_{[-, +, +]} - 1 \rr
\end{equation}

\subsection{$\mathbf{\{Y_1, W_1\}_{num}^{\prime}}$}
\label{11.num.prime}
This product vanishes if any factor satisfies

\begin{multline}
\label{11.num.prime.zeros}
E[ - a + b, W_1, Y_1, \square] + \mu - 2 \alpha_0 =
\\
- 
\frac{1}{2} \ll - m_a + m_b + m_{\mu} \rr \, \apos
-
\frac{1}{2} \ll - n_a + n_b + n_{\mu} \rr \, \aneg
+ A^{ }_{W_1}   \, \apos - L^{+}_{Y_1}    \, \aneg
= 0,
\end{multline}

\noindent which leads to the zero-conditions

\begin{equation}
\label{11.num.prime.2.conditions}
  A^{ }_{\square, W_1}   = M_{[-, +, +]}     + c\  p, 
\quad
- L^{ }_{\square, Y_1}   = N_{[-, +, +]} + 1 + c\ \pp 
\end{equation}

\noindent Following subsection {\bf \ref{11.num}}, we should set $c=0$, 
and (\ref{11.num.prime.2.conditions}) translates to the non-zero-condition 

\begin{equation}
\label{11.num.prime.1.condition}
Y^{\intercal}_{1, \ssigma} - W^{\intercal}_{1, \ssigma + M_{[-, +, +]}} \geq - N_{[-, +, +]} 
\end{equation}

\subsection{The remaining six products} 
\label{remaining.six.products}
The analysis of the remaining six products is identical to that in subsections {\bf \ref{11.num}} 
and {\bf \ref{11.num.prime}}, and it suffices to list the non-zero-condition in each case.

\subsubsection{$\mathbf{\{Y_2, W_2\}_{num}}$}

\begin{equation}
\label{22.num.1.condition}
W^{\intercal}_{2, \ssigma} - Y^{\intercal}_{2, \ssigma + [p - M_{[+, -, +]} - 1]} \geq - \ll \pp - N_{[+, -, +]} - 1 \rr
\end{equation}

\subsubsection{$\mathbf{\{Y_2, W_2\}_{num}^{\prime}}$}

\begin{equation}
\label{22.num.prime.1.condition}
Y^{\intercal}_{2, \ssigma} - W^{\intercal}_{2, \ssigma + M_{[+, -, +]}}           \geq             - N_{[+, -, +]} 
\end{equation}

\subsubsection{$\mathbf{\{Y_1, W_2\}_{num}}$}

\begin{equation}
\label{12.num.1.condition}
W^{\intercal}_{2, \ssigma} - Y^{\intercal}_{1, \ssigma + M_{[+, +, -]}}           \geq             - N_{[+, +, -]} 
\end{equation}

\subsubsection{$\mathbf{\{Y_1, W_2\}_{num}^{\prime}}$}
\begin{equation}
\label{12.num.prime.1.condition}
Y^{\intercal}_{1, \ssigma} - W^{\intercal}_{2, \ssigma + [p - M_{[+, +, -]} - 1]} \geq   - \ll \pp - N_{[+, +, -]} - 1 \rr
\end{equation}

\subsubsection{$\mathbf{\{Y_2, W_1\}_{num}}$}
\begin{equation}
\label{21.num.1.condition}
W^{\intercal}_{1, \ssigma} - Y^{\intercal}_{2, \ssigma + [p - M_{[+, +, +]} - 1] - 1} \geq 1 - \ll \pp - N_{[+, +, +]} - 1 \rr
\end{equation}

\subsubsection{$\mathbf{\{Y_2, W_1\}_{num}^{\prime}}$}
\begin{equation}
\label{21.num.prime.1.condition}
Y^{\intercal}_{2, \ssigma} - W^{\intercal}_{1, \ssigma + M_{[+, +, +]} + 1} \geq - \ll N_{[+, +, +]} + 1 \rr
\end{equation}

\subsubsection{Remark} Equations (\ref{12.num.prime.1.condition}) and (\ref{21.num.1.condition}) make sense as 
Burge-type conditions because of the bounds in (\ref{upper.bound.on.M}) and (\ref{upper.bound.on.N}).

\subsection{If the denominator is zero, then the numerator is zero}
\label{if.den.is.zero.then.num.is.zero}
The non-zero-conditions on the $\{Y_i, W_j\}_{num}$ and $\{Y_i, W_j\}_{num}^{\prime}$ 
products that appear in the numerator can be combined in pairs to produce 
non-zero-conditions on $\{Y_i, Y_j\}$ and $\{W_i, W_j\}$ pairs, $i\neq j$ pairs also 
in the numerator, that can be compared to the first and second non-zero-conditions 
(\ref{Burge.condition.01}) and 
(\ref{Burge.condition.02}) obtained from the denominator.


\subsubsection{$\mathbf{\{Y_1, W_1\}_{num}}$ and $\mathbf{\{Y_2, W_1\}^{\prime}_{num}}$} 
Consider conditions (\ref{11.num.1.condition}) and (\ref{21.num.prime.1.condition}). 
We eliminate $W^{\intercal}_{1}$ to obtain a non-zero condition on $Y_1$ and $Y_2$ 
by re-writing (\ref{11.num.1.condition}) as

\begin{equation}
W^{\intercal}_{1, \ssigma + [ M_{[+, +, +]} + 1 ]} - 
Y^{\intercal}_{1, \ssigma + [p - M_{[-, +, +]} - 1] + [ M_{[+, +, +]} + 1 ]} 
\geq 
- \pp + N_{[-, +, +]} + 1
\end{equation}

\noindent which combined with condition (\ref{21.num.prime.1.condition}) gives

\begin{equation}
\label{Y1W1W1Y2}
Y^{\intercal}_{2, \ssigma} - 
Y^{\intercal}_{1, \ssigma + [r_a - 1] + p} \geq 1 - s_a - \pp 
\end{equation}

\noindent which is a weak version of condition (\ref{Burge.condition.01}).



\subsubsection{$\mathbf{\{Y_1, W_2\}_{num}}$ and $\mathbf{\{Y_2, W_2\}^{\prime}_{num}}$} 
From (\ref{12.num.1.condition}) and (\ref{22.num.prime.1.condition}), 

\begin{equation}
\label{Y1W2W2Y2}
Y^{\intercal}_{2, \ssigma} - 
Y^{\intercal}_{1, \ssigma + [r_a - 1]} \geq 1 - s_a 
\end{equation}

\noindent which is condition (\ref{Burge.condition.01}).



\subsubsection{$\mathbf{\{Y_2, W_1\}_{num}}$ and $\mathbf{\{Y_1, W_1\}^{\prime}_{num}}$} 
From (\ref{21.num.1.condition}) and (\ref{11.num.prime.1.condition}), 

\begin{equation}
\label{Y2W1W1Y1}
Y^{\intercal}_{1, \ssigma} - 
Y^{\intercal}_{2, \ssigma + [p - r_a] - 1} 
\geq 
1 - [\pp - s_a],
\end{equation}

\noindent which is condition (\ref{Burge.condition.02}).



\subsubsection{$\mathbf{\{Y_2, W_2\}_{num}}$ and $\mathbf{\{Y_1, W_2\}^{\prime}_{num}}$} 
From (\ref{22.num.1.condition}) and (\ref{12.num.prime.1.condition}),

\begin{equation}
\label{Y2W2W2Y1}
Y^{\intercal}_{1, \ssigma} - 
Y^{\intercal}_{2, \ssigma + [p - r_a - 1] + p} \geq 1 - [\pp - s_a] - \pp,
\end{equation}

\noindent which is a weak version of condition (\ref{Burge.condition.02}). 



\subsubsection{$\mathbf{\{Y_1, W_1\}^{\prime}_{num}}$ and $\mathbf{\{Y_1, W_2\}_{num}}$} 
From (\ref{11.num.prime.1.condition}) and (\ref{12.num.1.condition}),

\begin{equation}
\label{W1Y1Y1W2}
W^{\intercal}_{2, \ssigma} - 
W^{\intercal}_{1, \ssigma + r_b - 1} \geq 1 - s_b,
\end{equation}

\noindent which is condition (\ref{Burge.condition.01}).



\subsubsection{$\mathbf{\{Y_2, W_1\}^{\prime}_{num}}$ and $\mathbf{\{Y_2, W_2\}_{num}}$} 
From (\ref{21.num.prime.1.condition}) and (\ref{22.num.1.condition}),

\begin{equation}
\label{W1Y2Y2W2}
W^{\intercal}_{2, \ssigma} - 
W^{\intercal}_{1, \ssigma + [r_b  - 1] + p} \geq 1 - s_b - \pp
\end{equation}

\noindent which is a weak version of condition (\ref{Burge.condition.01}).



\subsubsection{$\mathbf{\{Y_1, W_2\}^{\prime}_{num}}$ and $\mathbf{\{Y_1, W_1\}_{num}}$} 
From (\ref{12.num.prime.1.condition}) and (\ref{11.num.1.condition}),

\begin{equation}
\label{W2Y1Y1W1}
W^{\intercal}_{1, \ssigma} - 
W^{\intercal}_{2, \ssigma + [p - r_b - 1] + p} 
\geq 
1 - [\pp - s_b] - \pp
\end{equation}

\noindent which is a weak version of condition (\ref{Burge.condition.02}).



\subsubsection{$\mathbf{\{Y_2, W_2\}^{\prime}_{num}}$ and $\mathbf{\{Y_2, W_1\}_{num}}$} 
From (\ref{22.num.prime.1.condition}) and (\ref{21.num.1.condition}),

\begin{equation}
\label{W2Y2Y2W1}
W^{\intercal}_{1, \ssigma} - 
W^{\intercal}_{2, \ssigma + [p - r_b - 1]} 
\geq 
1 - [\pp - s_b] 
\end{equation}

\noindent which is condition (\ref{Burge.condition.02}).

The stronger condition in each of the above cases is one of the Burge conditions. Thus, when 
the denominator $z_{den}$ of the building block partition function $Z_{bb}$ is non-zero, then 
the numerator $z_{num}$ is also non-zero. The reverse is not true. 

Note that the above result is similar but different from that in \cite{zamolodchikov}, where 
Zamolodchikov argues that 
{\bf 1.} The conformal block $\cB^{\, \textit{gen}}_n$ is a meromorphic function in $\Delta_a$, 
the conformal dimension of the Virasoro irrep that flows in a channel, and that $\cB^{\, \textit{gen}}_n$
has only simple poles at $\Delta_a = \Delta_{a_{r, s}}$, where 
$a_{r,s} = - \frac{1}{2} [r \alpha_+ +s \alpha_-]$, and
{\bf 2.} If the fusion rules are satisfied, then the residue at each pole vanishes.

Our result is that 
{\bf 1.} When a summand in $Z_{bb}$ has a zero in the denominator, and the fusion rules are 
satisfied, then it also has a zero in the numerator. This is independent of Zamolodchikov's 
statement, since in the latter, the whole sum vanishes rather than just the summand with the 
zero in the denominator. 
{\bf 2.} Zamolodchikov has argued that  $\cB^{\, \textit{gen}}_n$ has only simple poles, 
while, as far as we can tell, summands in $Z_{bb}$ can have poles of order greater then 1.

\section{An Ising conformal block}
\label{ising.conformal.block}

In this section, we set $p=3$ and $\pp=4$, so that the minimal model component $\cM^{p, \pp}$ 
of the conformal field theory $\cM^{p, \pp, \cH}$ under consideration, is the Ising model.
In this case, there are three primary fields to form conformal blocks from. They can be 
labeled as follows. 
$\{r, s\} = \{1, 1\}$ is the identity operator $\mathbbm{1}$, 
$\{r, s\} = \{1, 2\}$ is the spin     operator $\sigma$ and 
$\{r, s\} = \{1, 3\}$  is the thermal operator $\psi$.
Explicit expressions for conformal blocks can be found in \cite{ardonne.sierra} and references 
therein. Consider the 6-point conformal block of $\sigma$ fields in Figure 
{\bf \ref{ising.conformal.block}}.

%
\begin{center}
\begin{minipage}{5.0in}
\setlength{\unitlength}{0.001cm}
\renewcommand{\dashlinestretch}{30}
\begin{picture}(4800, 3000)(-3000, -1000)
%
\thicklines
%
\path(0000,0000)(6000,0000)
\path(1200,0000)(1200,1200)
\path(2400,0000)(2400,1200)
\path(3600,0000)(3600,1200)
\path(4800,0000)(4800,1200)
\put(1700,-0400){$\mathbbm{1}$}
\put(2900,-0400){$\sigma$}
\put(4100,-0400){$\mathbbm{1}$}
\put(1100,1400){$\sigma$}
\put(2300,1400){$\sigma$}
\put(3500,1400){$\sigma$}
\put(4700,1400){$\sigma$}
\put(-0400,-0100){$\sigma$}
\put( 6200,-0100){$\sigma$}
\end{picture}
\begin{ca}
\label{ising.conformal.block.diagram}
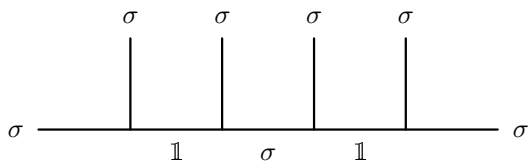
The comb diagram representation of the Ising $6$-point conformal block discussed in {\bf
\ref{ising.conformal.block}}.
All external lines correspond to vertex operator insertion of the spin operator $\sigma$.
The internal channels carry the Virasoro irrep's that correspond to the identity operator,
the spin operator, then the identity operator.
\end{ca}
\end{minipage}
\end{center}
\bigskip

In this case, 
$\apos     =   \sqrt{4/3}$, 
$\aneg     = - \sqrt{3/4}$, 
and
$\alpha_{1,2} = - \frac{1}{2} \aneg = - \frac{1}{2} \sqrt{3/4}$, and following 
\cite{ardonne.sierra}, 

\begin{multline}
\label{eq.sconfblock}
\langle \sigma(z_0) \sigma(z_1) \cdots \sigma(z_5) \rangle 
= 
\\
\frac{1}{2} 
\prod_{i=1}^3 
\ll z_{2i-2} - z_{2i-1} \rr^{-\frac{1}{8}}
\ll
\sum_{ \substack{t_1=1, \\ t_2,t_3=-1,1}}
\prod_{i=1}^3 t_i 
\prod_{1 \leq i < j \leq 3}
\ll 1-x_{i,j} \rr^{\frac{t_i t_j}{4}}
\rr^{\frac{1}{2}}, 
\end{multline}

\noindent where 

\begin{equation}
x_{i,j}=\frac{(z_{2i-2}-z_{2i-1})(z_{2j-2}-z_{2j-1})}{(z_{2i-2}-z_{2j-1})(z_{2j-2}-z_{2i-1})}
\end{equation}

\noindent setting the coordinates 
$z_0=0$, 
$z_1=q_1q_2q_3$, 
$z_2=q_2q_3$, 
$z_3=q_3$, 
$z_4=1$, 
$z_5=\infty$,

\begin{equation}
x_{1, 2} = \frac{-q_1(1-q_2)}{1-q_1},
\quad 
x_{1, 3} = q_1 q_2 q_3,
\quad 
x_{2, 3} = \frac{q_3(1-q_2)}{1-q_2q_3}.
\end{equation}

\noindent The instanton partition function should equal the product of the Ising conformal 
block and a contribution from the Heisenberg algebra $\cH$
\footnote{\ The contribution of the Heisenberg algebra $\cH$ is often referred to as 
{\it the $U(1)$ factor}.
}. 
Using {\it e.g.} \cite{alba.fateev.litvinov.tarnopolskiy}, equation (1.9), 

\begin{equation}
Z = 
\ll
\prod_{1 \leq i \leq j \leq 3} (1-q_i\cdots q_j)^{\frac{1}{8}}
\rr
\langle
\sigma(z_0) \sigma(z_1) \cdots \sigma(z_5)
\rangle
\end{equation}

\noindent Therefore 

\begin{multline}
Z =
1 - \frac{1}{8}  q_1     - \frac{1}{8}   q_3
  - \frac{5}{128}q_1^2   - \frac{1}{8}   q_1 q_2
  + \frac{1}{64} q_1 q_3 
\\  
  - \frac{5}{128} q_2^2
  - \frac{1}{8}  q_2 q_3 - \frac{5}{128} q_3^2
  + \cdots
  - \frac {453}{8192} q_1^2 q_2^2 q_3^2 + \cdots
\end{multline}

\noindent Calculating the expansion of $Z$ up to degree 2 in each variable, we find that 
result coincides with the sum of non-zero terms in the instanton partition function. Using 
the notation

\begin{multline}
Z_{Nek} \ll Y_1, Y_2 \ | \ Y_3, Y_4 \ | \ Y_5, Y_6 \rr
= 
\\
Z_{bb} \ll \varnothing, \varnothing \ | \ Y_1, Y_2 \rr \cdot 
Z_{bb} \ll Y_1, Y_2                 \ | \ Y_3, Y_4 \rr \cdot 
Z_{bb} \ll Y_3, Y_4                 \ | \ \varnothing, \varnothing \rr
\end{multline} 

\noindent the $q_1^2q_2^2q_3^2$-term, as an example, is 

\begin{multline}
Z_{Nek} \ll 2, \varnothing | \varnothing, 2 | 2, \varnothing \rr +
Z_{Nek} \ll 2, \varnothing |           1, 1 | 2, \varnothing \rr +
Z_{Nek} \ll 2, \varnothing | 2, \varnothing | 2, \varnothing \rr 
\\
+
Z_{Nek} \ll 2, \varnothing | 2, \varnothing | 1 + 1, \varnothing \rr +
Z_{Nek} \ll 2, \varnothing | 1, 1           | 1 + 1, \varnothing \rr =
- \frac{453}{8192}
\end{multline}

\noindent while all other terms, that satisfy Proposition \ref{proposition.01}
and the condition $|Y_1|+|Y_2|=|Y_3|+|Y_4|=|Y_5|+|Y_6| = 2$, vanish. 

\section{An explanation, based on a conjecture, of why we obtain $\cM^{p, \pp, \cH}$ 
conformal blocks} 
\label{why.we.get}

As mentioned in section {\bf \ref{introduction}}, there is a proof of AGT in the context 
of conformal blocks in $\cM^{\textit{gen}, \cH}$ with non-degenerate intermediate Virasoro 
representations in \cite{alba.fateev.litvinov.tarnopolskiy}. In this subsection, we use 
{\bf 1.} results from \cite{alba.fateev.litvinov.tarnopolskiy}, 
{\bf 2.} Proposition \ref{proposition.01} of section {\bf \ref{modified.denominator}}, 
{\bf 3.} that the generating function of Burge pairs is the character of 
$\cH^{p, \pp, \cH}_{r, s}$ \cite{foda.lee.welsh, feigin.feigin.jimbo.miwa.mukhin}, and
{\bf 4.} Conjecture  \ref{conjecture.01} below, 
to explain why restricting the summation to Burge pairs as in \eqref{restricted.sum} leads
to conformal blocks in $\cM^{p, \pp, \cH}$. Proving {\bf 3} would amount to proving that 
restricting to Burge pairs leads to conformal blocks in $\cM^{p, \pp, \cH}$, but this is 
beyond the scope of this work.

Consider the Verma module $\cH_a$ over $\cV^{gen} \oplus \cH$ generated by highest-weight 
vector $| a \rangle$,
\begin{equation}
L_k | a \rangle = a_k |a \rangle = 0, \ \ k > 0, \quad L_0 | a \rangle = \Delta_a |a \rangle,
\end{equation}
\noindent where $L_k$, and $a_k$, $k \in \mathbb{Z}$, are generators of $\cV^{gen}$ and 
$\cH$, respectively, and
\begin{equation}
\Delta_a = \frac{1}{4} \ll b_{gen} + \frac{1}{b_{gen}} \rr^2 - a^2, \quad 
b_{gen} = \ll \frac{\epsilon_2}{\epsilon_1} \rr^{\frac{1}{2}}. 
\end{equation}

Conformal blocks are defined in terms of vertex operators $\cO_{\mu}(z): \cH_a \mapsto \cH_b$,
that in turn are defined by the commutation relations
\begin{equation}
[L_k,\cO_{\mu}(z)] = z^{k+1} \partial_z \cO_{\mu}(z) + i (k+1) \Delta_{\mu^{\prime}} q^{k} \cO_{\mu}(q),
\quad
\mu^{\prime} = \mu - \frac{\epsilon_1+\epsilon_2}{2}
\end{equation}

\noindent as well as
\begin{equation}
[a_k, \cO_{\mu}(z)] =i    \mu  z^k \cO_{\mu}(z), \ \ k<0, 
\quad
[a_k, \cO_{\mu}(z)] =i (Q-\mu) z^k \cO_{\mu}(z), \ \ k>0
\end{equation}

AGT was proven in \cite{alba.fateev.litvinov.tarnopolskiy} for generic central charge $c_{\textit{gen}}$, 
in the following sense

\begin{Proposition} 
\label{proposition.02}
Following \cite{alba.fateev.litvinov.tarnopolskiy}, there exists an orthogonal 
basis $J_{\vec{Y}}$ $\in$ $\cH_a$ labeled by pairs of Young diagrams such that the matrix elements 
of vertex operator $\cO_{\mu}$ satisfy 

\begin{equation}
\label{eq:AFLT}
\frac{\langle J_{\vec{Y}}           \ | \ \cO_{\mu}\ | \ J_{\vec{W}} \rangle}{
      \langle J_{\vec{\varnothing}} \ | \ \cO_{\mu}\ | \ J_{\vec{\varnothing}} \rangle}
      =
      z_{num} \ll \vec a, \vec Y\  | \ \mu \ | \ \vec b, \vec W \rr 
\end{equation}

\noindent where $\vec{a}=\{a, -a\}$, $\vec{b}=\{b, -b\}$.
\end{Proposition}

\noindent From this proposition, it follows that 

\begin{equation}
\label{eq:norm}
\langle J_{\vec{Y}}\ |\ J_{\vec{W}} \rangle
=
z_{norm} \ll \vec a, \vec Y \ \rr \ \delta_{\vec{Y},\vec{W}} 
\end{equation}

The vectors $J_{\vec{Y}}$ can be written in the standard basis of the Verma module,

\begin{equation}\label{eq:JYinLa}
J_{\vec{Y}} =
\sum_{\lambda,\mu}
C_{\vec{Y}}^{\lambda,\mu} 
L_{-\lambda_1}L_{-\lambda_2}\cdots a_{-\mu_1}a_{-\mu_2} \cdots |a\rangle,
\end{equation}

\noindent where the summation is over partition pairs $\{\lambda,\mu\}$ such that 
$|\lambda|+|\mu|=|Y_1|+|Y_2|$. The coefficients $C_{\vec{Y}}^{\lambda,\mu}$ depend 
on the parameters $a, b_{gen}$. In \cite[Corollary 3.8]{alba.fateev.litvinov.tarnopolskiy}, 
it was proven that $C_{\vec{Y}}^{\lambda,\mu}$ is a polynomial in $a$. In this section, we 
need the following conjecture
\footnote{ \ 
This conjecture is not original to this work. It is standard in the community, although 
not written in the literature.
}

\begin{Conj}
\label{conjecture.01}
The coefficients $C_{\vec{Y}}^{\lambda,\mu}$ are Laurent polynomials in $b_{gen}$. 
\end{Conj}

Conjecture \ref{conjecture.01} is motivated by the explicit examples of the vectors $J_{\vec{Y}}$ 
\cite{alba.fateev.litvinov.tarnopolskiy}. Further motivation is provided by the relation between 
Jack symmetric functions $J_{Y}^{\alpha}$ \cite{Macdonald.book} and $J_{\vec{Y}}$, for 
$\alpha=b_{gen}^2$ \cite{alba.fateev.litvinov.tarnopolskiy}. Namely, from Macdonald's conjectures, 
proved by Haiman \cite{haiman}, the coefficients of $J_{Y}^{\alpha}$, in the standard basis, are 
polynomial in $\alpha$, so it is natural to expect that $J_{\vec{Y}}$ satisfies an analogous property.

Assuming Conjecture \ref{conjecture.01}, we can set $b_{gen} \rightarrow b_{p, \pp}$ and 
$a \rightarrow a_{r, s}$, as defined in subsection {\bf \ref{parameterisation.minimal.models}}. 
Thus, we can consider $J_{\vec{Y}}$ as vectors in the module over $\cH \oplus \mathcal{V}^{p, \pp}$.
In this case, the Verma module $\cH_a$ becomes reducible, the maximal submodule $\mathrm{Ker}_a$ 
is a kernel of the Shapovalov form $\langle \cdot |\cdot \rangle$ on $\cH_a$, and the irreducible 
quotient is $\cH^{p, \pp, \cH}_{r, s} = \cH^{\, p, \pp}_{r, s} \otimes \cF$.

It was follows from (\ref{modified.denominator}) and (\ref{eq:norm}) that $\vec Y$ satisfy 
condition (\ref{conditions.01.simpler}) if and only if the vector $J_{\vec{Y}}$ is not in 
the kernel $\mathrm{Ker}_a$. Then to each Burge pair $\vec{Y}_{\textit{Burge}}$ one can 
consider $J_{\vec{Y}}$ as an element of quotient $\cH^{p, \pp, \cH}_{r,s}$.
The vectors $J_{\vec Y}$ $\in$ $\cH^{p, \pp, \cH}_{r, s}$, for a Burge pair ${\vec Y}_{\textit{Burge}}$, 
are linearly independent since they are orthogonal. It was proven in 
\cite{foda.lee.welsh, feigin.feigin.jimbo.miwa.mukhin}, that the generating function of Burge pairs give 
the character of $\cH^{p, \pp, \cH}_{r, s}$. 
Therefore the vectors $J_{\vec{Y}_{\textit{Burge}}}$ form a basis in $\cH^{p, \pp, \cH}_{r, s}$. This is 
the point of this subsection.

Finally we note that the norms and matrix elements of $\cO_{\mu}$ depend on the parameters $\{a, b, 
\mu, \epsilon_1, \epsilon_2\}$ algebraically. Therefore, since (\ref{eq:AFLT}) was proven for a generic 
central charge, it holds for the $\cM^{p, \pp, \cH}$ models, and the expression that we obtain for 
$\cB^{p, \pp, \cH}_n$, by summing over Burge pairs, holds.

\section{Generic model conformal blocks with Degenerate intermediate representations}
\label{degenerate}

In generic models with a generic central charge $c_{gen}$, $z_{norm} [\vec a, \vec Y]$ can have zeros 
due to degenerate $\cV^{gen}$ representations in the intermediate channels. Since $\vec a = \{a, -a\}$, 
setting $2 a = 2 a_{r, s} = - r \beta_+ - s \beta_-$, we can study these zeros just as in section {\bf
\ref{modified.denominator}}. Since the central charge is generic we have 

\begin{equation}
\label{eq:twozero}
C_{+} \, \beta_+ + C_{-} \, \beta_- = 0, 
\end{equation}

\noindent if and only if $C_+=C_-=0$, and only the $\mathbf{\{Y_1, Y_2\}_{\it den}}$ factors can be zero. 
Proceeding from the two zero-conditions \eqref{eq:twozero}, we obtain 

\begin{Proposition}
\label{proposition.03}
$z_{norm} [\vec a, \vec Y]\neq 0$
if and only if
$Y^{\intercal}_{2, \ssigma} - Y^{\intercal}_{1, \ssigma + r -1} \geq 1 - s$
\end{Proposition}

\noindent The proof of Proposition {\bf \ref{proposition.03}} follows the same line of arguments as 
the proof of Proposition {\bf \ref{proposition.01}} in section {\bf \ref{modified.denominator}}.
From Proposition {\bf \ref{proposition.03}}, we obtain

\begin{Proposition}
\label{proposition.04}

\begin{equation}
\label{eq:Bgen_rs=ZNek}
\cB^{degen, \cH}_n =
\sum_{\vec Y^{1}, \cdots, \vec Y^{n}}^{\prime}
\prod_{\i = 1}^{n+1} q_{\i}^{| \vec Y^{\i } |} 
Z_{bb} \ll \vec a^{\i - 1}, \vec Y^{\i - 1}\ | \ \mu^{\i } \ | \ \vec a^{\i}, \vec Y^{\i}, \rr
\end{equation}

\noindent where $\cB^{degen, \cH}_n$ is an $n$-channel generic model conformal block, such that 
some of the channels carry degenerate intermediate representations, and $\sum^{\prime}$ indicates 
that, for channels that carry degenerate representations, the sum is restricted to partition pairs 
that satisfy Proposition {\bf \ref{proposition.03}}.
\end{Proposition}

The proof of Proposition {\bf \ref{proposition.04}} is based on the same line of arguments as in 
section {\bf \ref{why.we.get}} but {\it without} requiring a conjecture analogous to Conjecture 
{\bf \ref{conjecture.01}}. Indeed, since the coefficients of $J_{\vec{Y}}$ are polynomial in $a$, 
we can set $a = a_{r, s}$ in \eqref{eq:JYinLa}. 
The vectors $J_{\vec{Y}}$ for which $z_{norm} [\vec a, \vec Y]=0$, belong to the kernel of the 
Shapovalov form on the Verma module $\cH_a$. Let $\cH_{r,s}^{\textit{gen}}$ denote the irreducible 
quotient of $\cH_a$. 
The vectors $J_{\vec{Y}}$, where $\vec Y$ satisfy Proposition {\bf \ref{proposition.03}}, project 
to the module $\cH_{r,s}^{\textit{gen}}$. In \cite{feigin.feigin.jimbo.miwa.mukhin}, Feigin {\it 
et al.} proved that the generating function of the $\vec{Y}$ pairs that satisfy Proposition {\bf 
\ref{proposition.03}} is the character of $\cH_{r,s}^{\textit{gen}}$, therefore the corresponding 
vectors $J_{\vec{Y}}$ form a basis in $\cH_{r,s}^{\textit{gen}}$. Using \eqref{eq:AFLT} and 
\eqref{eq:norm}, we obtain the expression \eqref{eq:Bgen_rs=ZNek} for the conformal block for 
degenerate representations.

\section{Comments and remarks}
\label{comments.remarks}

\subsection{$q$-$\mathfrak{gl}_{\infty}$ Ding-Iohara} 
Let $\cE$ be the algebra called $q$-deformed $\mathfrak{gl}_\infty$ in 
\cite{feigin.feigin.jimbo.miwa.mukhin}, and Ding-Iohara in 
\cite{feigin.hoshino.shibahara.shiraishi.yanagida} 
\footnote{\ Also called {\it \lq quantum toroidal $\mathfrak{gl}(1)$ algebra\rq}, 
                        {\it \lq elliptic Hall algebra\rq}, {\it etc.} in the literature.}. 
Following \cite{feigin.hoshino.shibahara.shiraishi.yanagida}, operators in the rank-2 
representations $\cF_{u_1}$ $\otimes$ $\cF_{u_2}$ of $\cE$ generate the sum of a $q$-deformed 
Virasoro algebra and a $q$-deformed Heisenberg algebra. 
On the other hand, following \cite{feigin.feigin.jimbo.miwa.mukhin} [Theorem 3.8], for special 
values of parameters $u_1$ and $u_2$, as well as $q_1$ and $q_3$ of $\cE$, this representation 
has a sub-representation with a basis labeled by Burge pairs. In is natural to expect that in 
the limit $q_1, q_3 \rightarrow 1$, the basis constructed in 
\cite{feigin.feigin.jimbo.miwa.mukhin} reduces to the basis $J_{\vec{Y}}$ described in 
section {\bf 7}.

\subsection{Higher-rank AGT-W} 
\label{agt-w}
AGT was extended to theories based on the higher rank algebras 
$\cW_N$ $\oplus$ $\cH$, $N > 2$, by Wyllard in \cite{wyllard}, and by 
Mironov and Morozov in \cite{mironov.morozov}. In this note, we chose to 
simplify the presentation by focusing on Virasoro minimal models, but we expect that our 
analysis extends without essential modification to minimal models based on $\cW_N$ algebras 
with $N>2$. We conjecture that the restricted partitions that are relevant to these extended 
cases are those that appeared in \cite{feigin.feigin.jimbo.miwa.mukhin}. 

\subsection{The work of Alkalaev and Belavin} In \cite{alkalaev.belavin}, Alkalaev and Belavin 
independently suggested the Virasoro result in (\ref{restricted.sum}) in the 4-point conformal 
block case. They proved a proposition equivalent to \ref{proposition.01}, made the same comment 
on conformal blocks in generic models with degenerate intermediate representations as in section 
{\bf \ref{degenerate}}, albeit without proving an analogue of Proposition {\bf \ref{proposition.04}}
and made the same $W_N$ conjecture as in subsection {\bf \ref{agt-w}}.

\subsection{Previous works on AGT in minimal models} There are two previous works on AGT in 
minimal models that we are aware of.
In \cite{santachiara.2}, Santachiara and Tanzini identify Moore-Read wave functions, which are 
minimal model conformal blocks of $\{r, s\} = \{1, 2\}$ and $\{2, 1\}$ vertex operators, with 
Nekrasov instanton partition functions, AGT is applied without modification to these conformal 
blocks and ill-defined expressions are made well-defined using a deformation scheme, as 
outlined in subsection {\bf \ref{deformations}}.
In \cite{santachiara.4}, Estienne, Pasquier, Santachiara and Serban interpret $W_n \oplus \cH$ 
minimal model conformal blocks of $\{r, s\} = \{1, 2\}$ and $\{2, 1\}$ vertex operators as 
wave functions of a trigonometric Calogero-Sutherland models with non-trivial braiding 
properties, and find that the excited states are characterized by $(n+1)$-partitions, 
just as in AGT. While Estienne {\it et al.} use different notation from ours, preliminary 
checks indicate that their partitions can be translated to the Burge pairs used in this note, 
for $n=2$, and $\{r, s\} = \{1, 2\}$ or $\{2, 1\}$. 

\subsection{Geometry} 
Let $\cM(r, N)$ be the moduli space of $U(r)$ instantons on $\mathbb{R}^4$. The instanton partition 
function for $\prod_{i=1}^n U(2)_{\i}$ gauge theory equals the generating function of equivariant 
integrals over $\cM (2, N_1) \times \cdots \times \cM (2, N_n)$, where the equivariant integral is 
taken with respect to the torus 
$\mathbb{T} =
(\mathbb{C}^*)^{2} \times (\mathbb{C}^*)^{2}_{1} \times 
(\mathbb{C}^*)^{2}_{2} \times \cdots \times 
(\mathbb{C}^*)^{2}_{n}$, where the first $(\mathbb{C}^*)^{2}$ acts on $\mathbb{C}^2$, and 
$(\mathbb{C}^*)^{2}_{i}$ acts on the $i$-th instanton moduli space $\cM (2, N_i)$ by 
constant gauge transformation. These equvariant integrals are computed using localization 
and are given by the sum over torus fixed points. These points were labeled by $n$ pairs 
of Young diagrams $\vec{Y}^1, \cdots, \vec{Y}^n$. 
The parameters $\epsilon_1,$ $\epsilon_2,$ and $\vec{a}^i$ are the coordinates on 
$\mathrm{t} = \mathrm{Lie}(\mathbb{T})$. In the $\cM^{p, \pp, \cH}$ case, $\epsilon_1$ and 
$\epsilon_2$ are linearly-dependent, and $a^i_j$ is given by (\ref{parameters.03}). Geometrically, 
this means that we are considering the one-dimensional subgroup
$\mathbb{C}^*_{\epsilon_1, \epsilon_2, \vec{a}^i}$ $\subset$ $\mathbb{T}$.

The function $z_{norm} [\vec a, \vec Y ]$ is the determinant of the vector field with coordinates 
$\{ \epsilon_1, \epsilon_2, $ $ a_1, a_2 \}$ on the tangent space of the point labeled by $\vec{Y}$. 
The condition $z_{norm} [\vec a, \vec Y] \neq 0$ is equivalent to the fact that corresponding point 
is an isolated fixed point of the one dimensional torus $\mathbb{C}^*_{\epsilon_1,\epsilon_2,\vec{a}^i}$. 
Therefore, summing over Burge pairs is equivalent to summing over the \emph{isolated} torus fixed points.

\section*{Acknowledgements}
We wish to thank the Simons Center for Geometry and Physics for hospitality where this work was started, 
Dr S Corteel and {\it Laboratoire d'Informatique Algorithmique: Fondements et Applications}, Paris 7, 
for hosting OF while it was completed.
MB was supported in part by RFBR grants {\tt 12-02-01092-a}, {\tt 13-01-90614}, 
and OF was supported by Australian Research Council and the {\it Fondation Sciences Math\'{e}matiques 
de Paris}. We wish to thank 
V Fateev,
B Feigin, 
I Kostov,
R Santachiara, 
D Serban and A Tanzini for discussions and useful remarks.

\end{document}